\RequirePackage{fix-cm}
\documentclass[twocolumn,epjc3]{svjour3}
\smartqed

\usepackage[english]{babel}
\usepackage{subcaption}   
\usepackage{graphicx}
\usepackage{amsmath, amssymb}
\usepackage{dcolumn}      
\usepackage{bm}           
\usepackage{hyperref}     

\journalname{Eur. Phys. J. C}

\begin{document}

\title{Long-Lived Oscillons and Domain Walls in the \texorpdfstring{$\mathbb{Z}_{2}$}{Z2}-Symmetric Two-Higgs-Doublet Model}
\author{Zhaoyu Meng \thanksref{e1}\thanksref{e2}}
\thankstext{e1}{e-mail: robert\_meng2020@163.com}
\thankstext{e2}{Current affiliation: Southern University of Science and Technology, Shenzhen, China. Current e-mail: 12631063@mail.sustech.edu.cn}
\institute{Huazhong University of Science and Technology}

\titlerunning{Long-Lived Oscillons and Domain Walls in the \texorpdfstring{$\mathbb{Z}_{2}$}{Z2}-Symmetric Two-Higgs-Doublet Model}
\authorrunning{Z. Meng}

\date{Received: date / Accepted: date} 

\maketitle

\begin{abstract}
Domain walls arising from spontaneous symmetry breaking in the early universe are typically unstable and decay rapidly. However, numerical simulations of the \texorpdfstring{$\mathbb{Z}_{2}$}{Z2}-symmetric two-Higgs-doublet model (2HDM) reveal an unexpected long-lived closed domain wall structure that emerges at late stages of decay, which we identify as a high-amplitude oscillon. Their longevity stems from a potential with two distinct vacua: an interior oscillating in one vacuum and an exterior static in the other. Within a specific parameter regime, radiation is highly suppressed and the lifetime can exceed cosmological timescales. To gain analytical insight into the mechanism, we first consider reduced models with fewer fields (2 and 4 real scalars). These are not approximations to the 2HDM but rather toy models that capture the essential feature of a \texorpdfstring{$\mathbb{Z}_{2}$}{Z2}-symmetric potential with two vacua. Floquet analysis further confirms the stability of these structures under perturbations in both the 4-field toy model and the full 2HDM.

\end{abstract}

\section{Introduction}

Topological defects form during spontaneous symmetry breaking when fields in different regions of space independently fall into one of the distinguishable vacuum minima as the system gradually cools down from the Big Bang \cite{PhysRevD.37.3052,Kibble_1976,Zeldovich:1974uw}. When the early universe gradually cools, defects like cosmic strings and domain walls can form \cite{Battye_2020,particle-interact,benabou2023signaturesprimordialenergyinjection,PhysRevLett.48.1867}, where their structures are key to investigating how the universe evolved during its earliest period \cite{particle-interact}. Domain walls are an important topological concept and could lead to various cosmological phenomena in the early universe, such as the Primordial black hole \cite{kitajima2025primordialblackholeformation}. There are various types of domain walls \cite{PhysRevLett.48.1867}, and this paper focuses on the wall arising from ${\mathbb{Z}_2}$ symmetry.

Closed domain wall structures tend to contract and decay due to their curvature, and collapse in an increasing rate. However, in this work, we demonstrate that at the end of the decay of such closed structure, the decay rate of the wall may diverge from the previous prediction, and the structure becomes a long-lived, closed localized domain wall. Some studies have shown that certain structures can resist decay, such as biased vacuum \cite{Dvali_1995} and kinky Vortons \cite{BATTYE2025139311}. In this work, we introduce a new explanation that these long-lived remnants are high-amplitude oscillons (specifically, Q-balls) that can become highly radiation-suppressed under a specific parameter choice.

There are many studies on oscillons in cosmology \cite{PhysRevLett58747,Amin_2010,Gleiser_1994,Copeland_1995}, but unlike most of them, which continuously radiate energy \cite{PhysRevLett58747,Amin_2010} and possess a preferred radius \cite{Gleiser_1994}, the large-amplitude oscillons identified here interpolate between two degenerate vacua and form self-trapped spherical domain walls. They are stable across the entire continuous range of allowed radii, with no preferred or unstable size, and can be fully radiation-suppressed when under some specific parameter choice. Our goal is to characterize this new class of long-lived bubble-like structures and to demonstrate their emergence from domain wall decay in the 2HDM.

To isolate the essential mechanism of radiation‑suppressed oscillons, we first study two simplified models: a two‑real‑scalar model and a two‑complex‑scalar (four‑real‑scalar) model. These models are not obtained by truncating the 2HDM; they are independent field theories that capture the core nonlinear dynamics without the complexity of gauge fields. Their purpose is to reveal the parametric condition for radiation suppression. Only after establishing the oscillon solution in these toy models do we embed it into the full 2HDM, respecting all gauge degrees of freedom.

\section{\texorpdfstring{$\mathbb{Z}_{2}$}{Z2}-symmetry 2-Higgs Doublet Model}

The 2-Higgs Doublet Model (2HDM) is a Beyond the Standard Model theory that combines two Higgs fields as a whole \cite{Battye_2020}. Compared to the single Higgs model which is constructed by two complex scalar fields, the 2HDM is composed of two complex doublets, and exhibits the Standard Model symmetries of $SU (2)\otimes U (1)$ as the single Higgs field does \cite{GABRIEL2007141,Battye_2011}, but the additional field involved allows the 2HDM to have more geometric characteristics due to the increased complexity of the vacuum manifold’s homotopy group \cite{GABRIEL2007141,Battye_2011,battye2024spontaneoushopffibrationtwohiggsdoublet,Sassi:2024dw}. By introducing the new properties as extensions to the basic Standard Model symmetry, 2HDM potentially offers new insights into unresolved questions in Standard Model and early universe physics areas, like supersymmetry \cite{Branco_2012}, CP-violation \cite{CP}, and dark matter \cite{dark}. 

Apart from basic Standard Model symmetries, some additional constraints can be applied to the 2HDM \cite{Battye_2021}. ${\mathbb{Z}_2}$ symmetry is one of the most commonly used constraints that allows the domain walls of the topological defects to be produced when applied \cite{Battye_2011,Battye_2020,Battye_2021}. Additionally, compared to alternative constraints such as CP1 and CP2 symmetries, ${\mathbb{Z}_2}$ symmetry makes both disjoint vacuum manifolds lie in the real space regime \cite{Battye_2011,Battye_2021,Sassi:2024dw}. In the ${\mathbb{Z}_2}$-2HDM, both the potential and kinetic parts of the Lagrangian are invariant under the transformation of $\Psi_1\rightarrow\Psi_1$ and $\Psi_2\rightarrow-\Psi_2$, where $\Psi_1$ and $\Psi_2$ are two complex scalar fields doublet in the 2HDM \cite{Battye_2020}. For 2HDM to exhibit such symmetry, some constraints need to be applied to the potential part of the Lagrangian \cite{Battye_2020,PhysRevD.76.039902,BATTYE2025139311,Sassi:2024dw}, where the Lagrangian under such constraint is \cite{Battye_2020,BATTYE2025139311,Sassi:2024dw}
\begin{equation}
\mathcal{L} \supset \left (\partial_\mu \Psi_1\right)^\dagger \left (\partial^\mu \Psi_1\right) +\left (\partial_\mu \Psi_2\right)^\dagger \left (\partial^\mu \Psi_2\right) -V (\Psi_1,\Psi_2),
\label{eq:lagrangian}
\end{equation}
however, QCD effects are beyond the scope of this work. The potential part of the Lagrangian is given by \cite{Battye_2011,PhysRevD.76.039902,BATTYE2025139311,Sassi:2024dw}.
\begin{equation}
\begin{aligned}
    V (\Psi_1, \Psi_2) =& -{\mu_1^2} (\Psi_1^\dagger \Psi_1) - {\mu_2^2} (\Psi_2^\dagger \Psi_2)\\& +{\lambda_1} (\Psi_1^\dagger \Psi_1)^2 + {\lambda_2} (\Psi_2^\dagger \Psi_2)^2 \\&+ {\lambda_3} (\Psi_1^\dagger \Psi_1) (\Psi_2^\dagger \Psi_2)\\& + { (\lambda_4 - |\lambda_5|)} \left(\mathrm{Re} (\Psi_1^\dagger \Psi_2)\right)^2 \\&+ { (\lambda_4 + |\lambda_5|)} \left(\mathrm{Im} (\Psi_1^\dagger \Psi_2)\right)^2,
    \label{eq:potential}
\end{aligned}
\end{equation}

and under such ${\mathbb{Z}_2}$, we may introduce a parameter $R^{1}$ to distinguish two vacuum manifolds, where the expression is given by \cite{Battye_2011,PhysRevD.76.039902,BATTYE2025139311}
\begin{equation}
    R^{1}=\Psi_1^\dagger \Psi_2 + \Psi_2^\dagger \Psi_1,
\end{equation}
where positive and negative signs of $R^{1}$ represent two distinct manifolds, and $R^{1}=0$ is the boundary between the two regions, which is called the domain wall. \cite{Battye_2021}.

In the detailed analysis and plotting of two complex doublet fields in the 2HDM system, we may break and decompose the system into 8 independent scalar fields. Fields in the 2-Higgs Doublet Models can be expressed as the combination of 8 scalar fields \cite{Battye_2021}:
\begin{equation}
    \Psi_1=\begin{pmatrix}\psi_1+i\psi_{3}\\ \psi_5+i\psi_{7}\end{pmatrix},\quad\quad\Psi_{2}=\begin{pmatrix}\psi_{2}+i\psi_4\\ \psi_{6}+i\psi_8\end{pmatrix},
\end{equation}
which can be used to expand the potential parts of the 2HDM Lagrangian as \cite{Battye_2021}
\begin{equation}
\begin{aligned}
    V=&-{\mu_1} (\psi_1^2+\psi_{3}^2+\psi_5^2+\psi_{7}^2)-{\mu_2} (\psi_{2}^2+\psi_4^2+\psi_{6}^2+\psi_8^2)\\&+{\lambda_1} (\psi_1^2+\psi_{3}^2+\psi_5^2+\psi_{7}^2)^2+{\lambda_2} (\psi_{2}^2+\psi_4^2+\psi_{6}^2+\psi_8^2)^2\\&+{\lambda_3} (\psi_1^2+\psi_{3}^2+\psi_5^2+\psi_{7}^2) (\psi_{2}^2+\psi_4^2+\psi_{6}^2+\psi_8^2)\\&+ (\lambda_4-|\lambda_5|) (\psi_1\psi_{2}+\psi_{3}\psi_4+\psi_5\psi_{6}+\psi_{7}\psi_8)^2\\&+ (\lambda_4+|\lambda_5|) (\psi_1\psi_4-\psi_{3}\psi_{2}+\psi_5\psi_8-\psi_{7}\psi_{6})^2,
\end{aligned}
\end{equation}
and the equations of motion for each of the $\psi_i$ are therefore given by \cite{Battye_2021}:
 
\begin{equation}
    \begin{aligned}
    \ddot{\psi}_{1,3,5,7}=&\nabla^2\psi_{1,3,5,7}+{\mu_1}\psi_{1,3,5,7}\\&-2{\lambda_1}\psi_{1,3,5,7} (\psi_1^2+\psi_{3}^2+\psi_5^2+\psi_{7}^2)\\&-\lambda_3\psi_{1,3,5,7} (\psi_{2}^2+\psi_4^2+\psi_{6}^2+\psi_8^2)\\&- (\lambda_4-|\lambda_5|)\psi_{2,4,6,8} (\psi_1\psi_{2}+\psi_{3}\psi_4\\&\qquad\qquad\qquad\qquad\qquad+\psi_5\psi_{6}+\psi_{7}\psi_8)\\&- (\lambda_4+|\lambda_5|)\psi_{4,8,-2,-6} (\psi_1\psi_4-\psi_{3}\psi_{2}\\&\qquad\qquad\qquad\qquad\qquad+\psi_5\psi_8-\psi_{7}\psi_{6}),
    \label{eq:eom1}
    \end{aligned}
\end{equation}
\begin{equation}
    \begin{aligned}
    \ddot{\psi}_{2,4,6,8}=&\nabla^2\psi_{2,4,6,8}+{\mu_2}\psi_{2,4,6,8}\\&-2{\lambda_2}\psi_{2,4,6,8} (\psi_{2}^2+\psi_4^2+\psi_{6}^2+\psi_8^2)\\&-\lambda_3\psi_{2,4,6,8} (\psi_1^2+\psi_{3}^2+\psi_5^2+\psi_{7}^2)\\&- (\lambda_4-|\lambda_5|)\psi_{1,3,5,7} (\psi_1\psi_{2}+\psi_{3}\psi_4\\&\qquad\qquad\qquad\qquad\qquad+\psi_5\psi_{6}+\psi_{7}\psi_8)\\&- (\lambda_4+|\lambda_5|)\psi_{-3,-7,1,5} (\psi_1\psi_4-\psi_{3}\psi_{2}\\&\qquad\qquad\qquad\qquad\qquad+\psi_5\psi_8-\psi_{7}\psi_{6}),
    \label{eq:eom2}
    \end{aligned}
\end{equation}

where we have introduced the notation $\psi_{-i}=-\psi_i$.

To investigate the 2HDM, one may approach it from simplified models that contain fewer degrees of freedom while retaining the same fundamental characteristics. For such toy models with simplified field theories, we may replace $\Psi_1$, $\Psi_2$ in the equations of motion with real scalar fields or complex scalar fields, corresponding to the 2-field model and the 4-field model. Using the 2-field model and 4-field model provides a simplified approach to 2HDM dynamics, as they involve a smaller number of evolving fields, making computation easier \cite{BATTYE2025139311}. Comparing the results from these different systems allows us to understand the effects of additional degrees of freedom.

\section{Simulation Setup and Methodology}
\subsection{Units and Dimensionless Parameters}
Throughout this work, we adopt natural units and express all quantities in units of the Higgs mass $M_h = 125\ \mathrm{GeV}$. Specifically, we set $M_h = 1$ so that masses, energies, and inverse lengths are measured in units of $M_h$. Consequently, space and time are measured in units of $1/M_h$ and $1/ (c M_h)$, respectively. Coupling constants are rendered dimensionless by appropriate powers of $M_h$; for instance, the quartic couplings $\lambda_i$ in the potential are already dimensionless in the Lagrangian, while the mass parameters $\mu_i^2$ are scaled as $\mu_i^2/M_h^2$.

All numerical values reported in the text are dimensionless. The physical values can be recovered by multiplying by the appropriate power of $M_h$: for example, the oscillon frequency in physical units is $\omega_{\text{phys}} = \omega \cdot M_h$, and the radial coordinate $r_{\text{phys}} = r / M_h$. This normalization simplifies the equations of motion and allows a direct comparison of different parameter regimes.
\subsection{System Setup}
By discretizing space into a grid, one can calculate the first and second derivative values of grids by the values of the fields of a point and its neighborhood. With parameters given in ~\ref{app:2hdm}, all grids are iterated in each time interval concurrently via the equation of motion given by equations (\ref{eq:eom1}) and (\ref{eq:eom2}) in the previous part, for time to progress in the system. We use the RK4 method in calculating our spatial derivative terms, with the system in flat space.

If the system starts with random initial conditions, each grid's initial values for evolving fields are set with random numbers that are uniformly distributed between -1 and +1, to simulate an early hot, dense, homogeneous universe and how it evolves when it starts to cool down \cite{BATTYE2025139311}. A damping term is added to the system during the initial iteration steps to simulate cooling \cite{BATTYE2025139311}. We may also apply continuous boundary conditions, which set in a way that the system is wrapped as a torus, by joining one side to its opposite. Periodic boundary conditions permit outgoing particles (or waves) to re‑enter the computational domain from the opposite side, and are suitable for modeling the early radiation-dense universe with highly frequent random collisions. But we should be aware that such boundary conditions may cause nonphysical wrapping strip mode \cite{BATTYE2025139311} at the end of wall evolution, and must be neglected in physical analysis.

To ensure that the observed long-term stability is not an artifact of wave reflection in approximating the infinite and non‑reflecting exterior, we implement an absorbing boundary condition (sponge layer) at the edges of the computational domain \cite{sponge}. This configuration effectively damps outgoing radiation and prevents the re-entry of emitted scalar waves, ensuring that the measured energy loss rates and life-times reflect the true physical decay of the oscillons into the vacuum. This approach involves introducing an artificial damping term into the equations of motion within a specified boundary region, where the damping intensity increases monotonically toward the edge of the domain. We define a coordinate-dependent damping coefficient $F (r)$ with a quadratic profile:
\begin{equation}
F (r) = 
\begin{cases} 
0 & \text{if } r \le R_\text{start} \\
5  \left ( \frac{r - R_{\text{start}}}{R_\text{max} - R_\text{start}} \right)^2 & \text{if } r > R_\text{start}
\end{cases}
\end{equation}
We set $R_{\text{start}} = 0.9 R_{\text{max}}$ as the threshold radius where the damping is initiated, where $R_{\text{max}}=32$ is used in this paper. To minimize numerical artifacts arising from oblique incidence at the boundaries, the sponge layer is implemented with a circular geometry, ensuring that outgoing waves strike the dissipative region at a near-normal angle. The evolving system includes all lattice grids with grid size $dx=0.25$ enclosed by circle with diameter $64$ (see ~\ref{app:convergence} for convergence test). In the equations of motion, this friction term manifests as an additional dissipative acceleration component: $\ddot{\psi}_{\text{damp}} = -F (r) \dot{\psi}$.

The numerical methods described above allow us to explore domain wall dynamics in the full 2HDM. Before analyzing the fundamental mechanism underlying the long-lived structures, we first present the results of those large-scale simulations. Then, we consider a minimal model that captures the essential physics while remaining analytically tractable.

The spatial and temporal convergence of the numerical scheme is verified in ~\ref{app:convergence}, ensuring that the observed long-lived behavior is not a numerical artifact. The source code and data necessary to reproduce all figures are available at the GitHub repository provided in ~\ref{app:git}.

\section{Simulation of Natural Domain Wall Decay}

\label{sec:natural}

\subsection{Late-time Evolution of Domain Walls Decay}

For the domain walls arising from spontaneous symmetry breaking in the early universe, we use random initial conditions and periodic boundary conditions to set up the system and observe how the structure of the wall changes when iterating in time. After the damping period in forcing the system to cool down, the domain walls will tend to contract to minimize its length as the wall has a higher energy density than the vacuum. Simulations show that the size of the domain wall structure (circumference) maintained non-zero values for the domain wall length for extended periods before ultimately collapsing into a single domain, as discussed above.

\begin{figure*}
    \centering
    \begin{subfigure}[b]{0.48\textwidth}
        \centering
        \includegraphics[width=\linewidth]{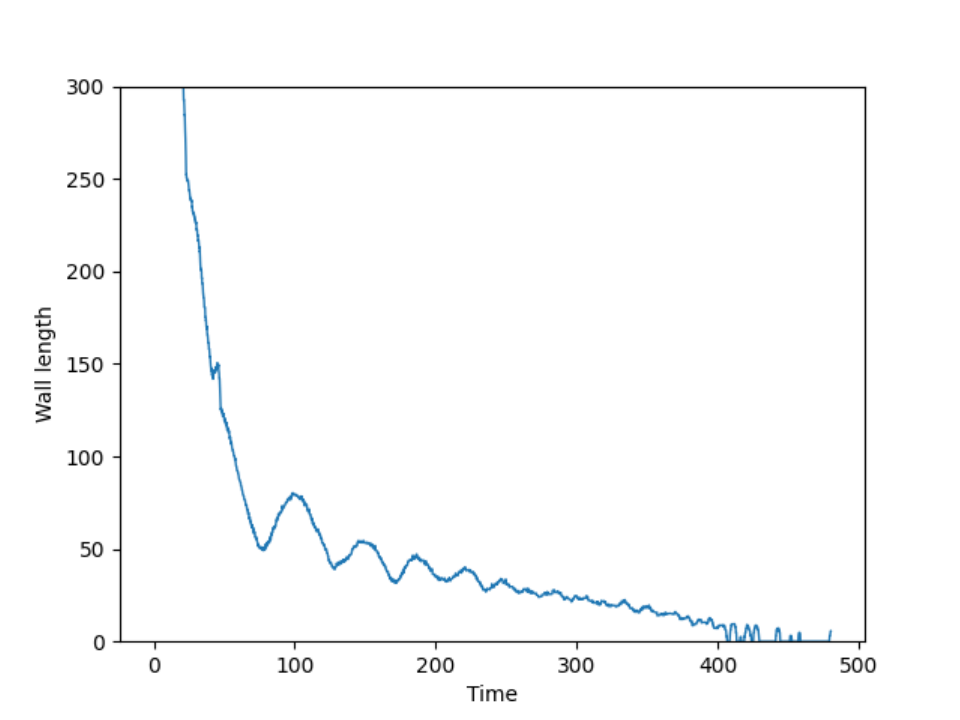}
        \caption{}
        \label{fig:wall_2d}
    \end{subfigure}
    \hfill
    \begin{subfigure}[b]{0.48\textwidth}
        \centering
        \includegraphics[width=\linewidth]{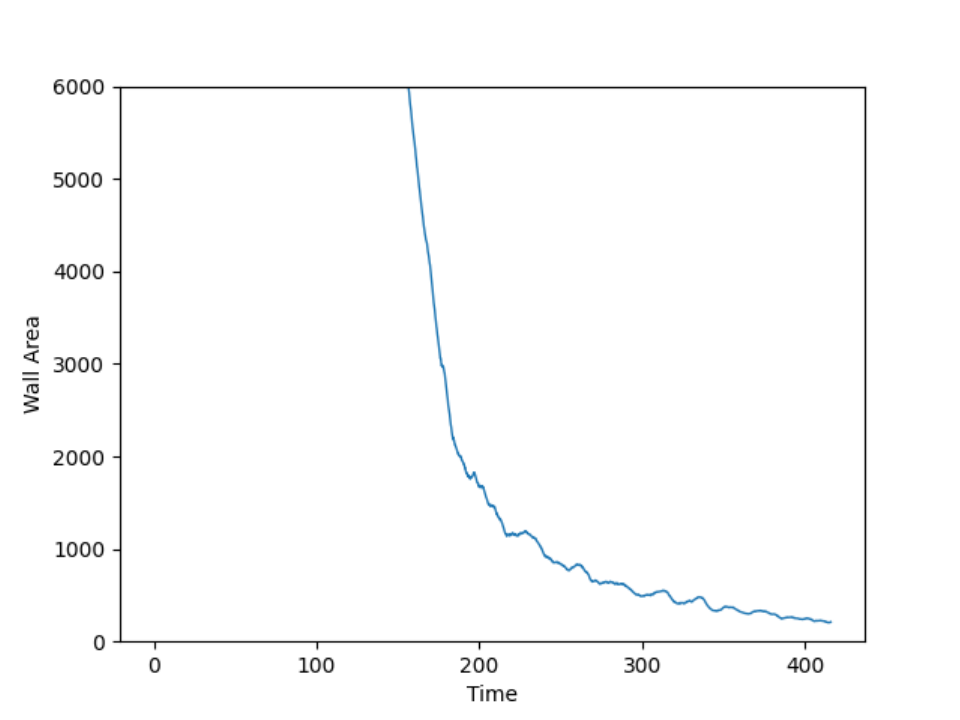}
        \caption{}
        \label{fig:wall_3d}
    \end{subfigure}

    \vspace{1em} 

    \begin{subfigure}[b]{0.32\textwidth}
        \centering
        \includegraphics[width=\linewidth]{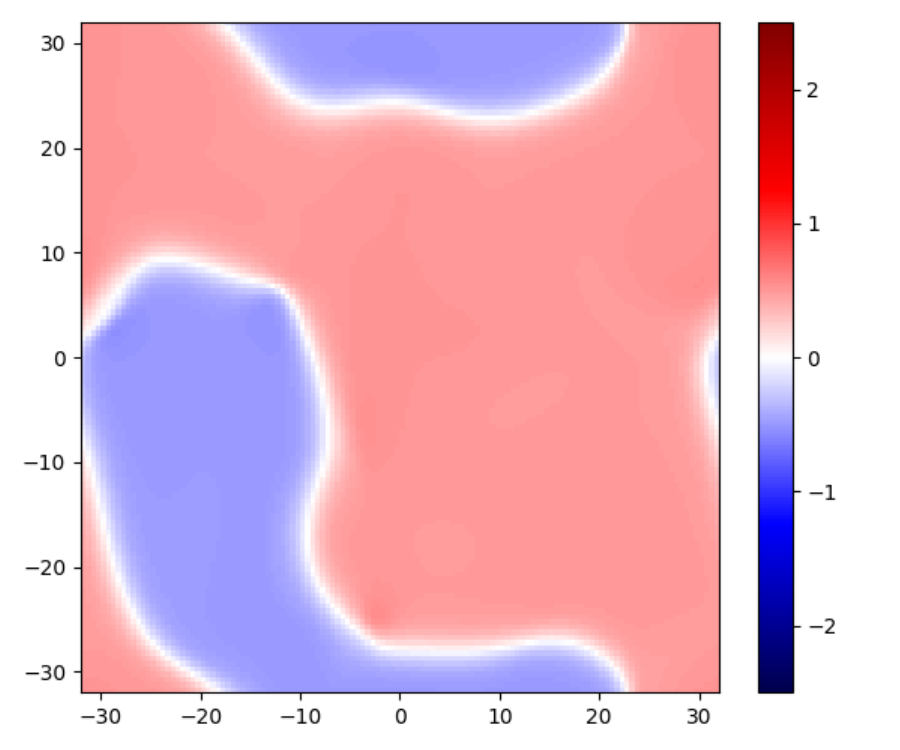}
        \caption{}
        \label{fig:2d_400}
    \end{subfigure}
    \hfill
    \begin{subfigure}[b]{0.32\textwidth}
        \centering
        \includegraphics[width=\linewidth]{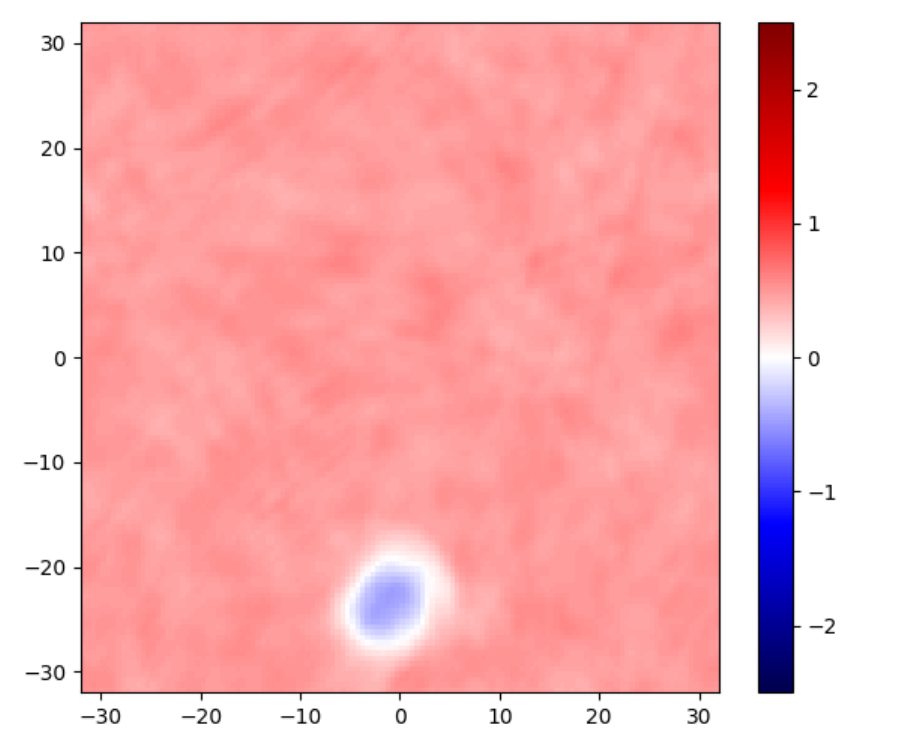}
        \caption{}
        \label{fig:2d_2000}
    \end{subfigure}
    \hfill
    \begin{subfigure}[b]{0.32\textwidth}
        \centering
        \includegraphics[width=\linewidth]{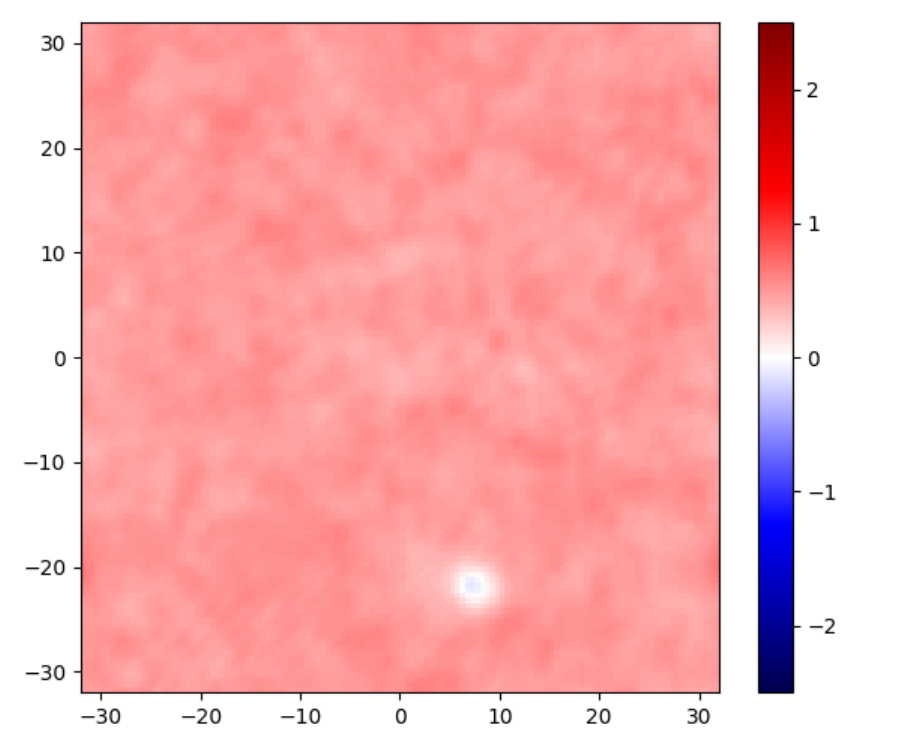}
        \caption{}
        \label{fig:2d_2800}
    \end{subfigure}

    \vspace{1em}

    \begin{subfigure}[b]{0.32\textwidth}
        \centering
        \includegraphics[width=\linewidth]{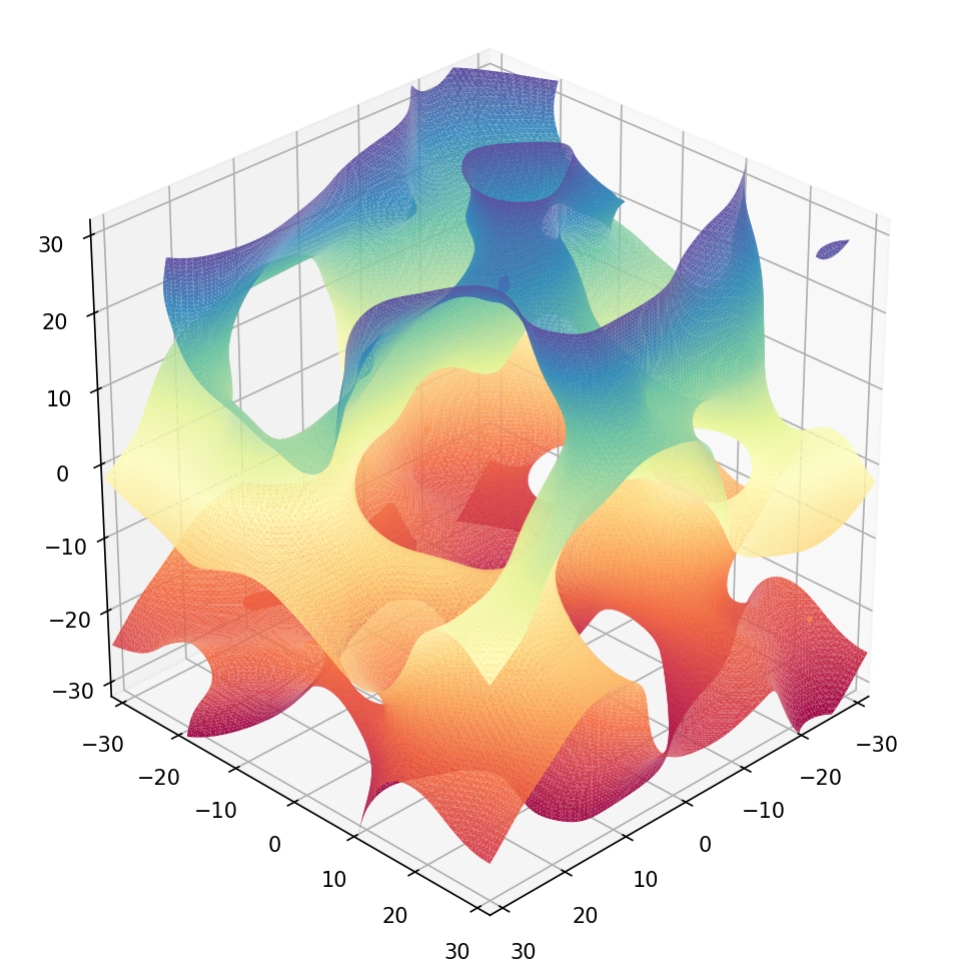}
        \caption{}
        \label{fig:3d_400}
    \end{subfigure}
    \hfill
    \begin{subfigure}[b]{0.32\textwidth}
        \centering
        \includegraphics[width=\linewidth]{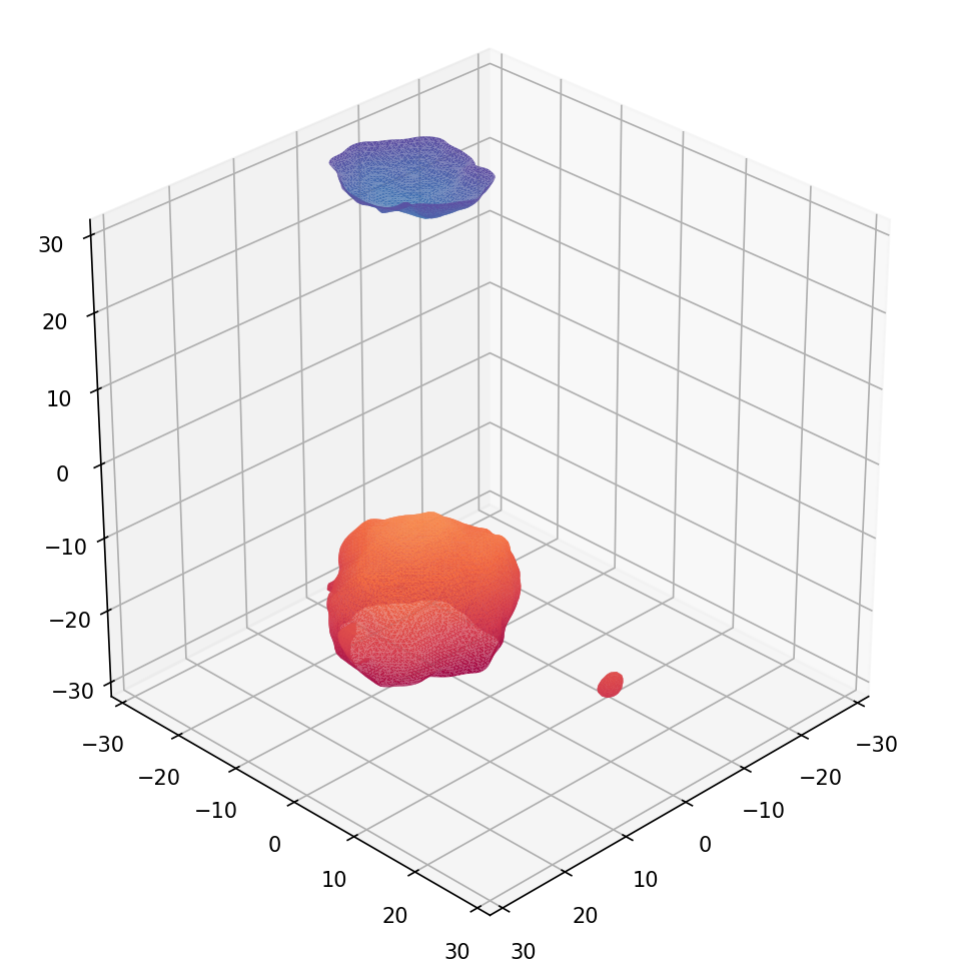}
        \caption{}
        \label{fig:3d_2000}
    \end{subfigure}
    \hfill
    \begin{subfigure}[b]{0.32\textwidth}
        \centering
        \includegraphics[width=\linewidth]{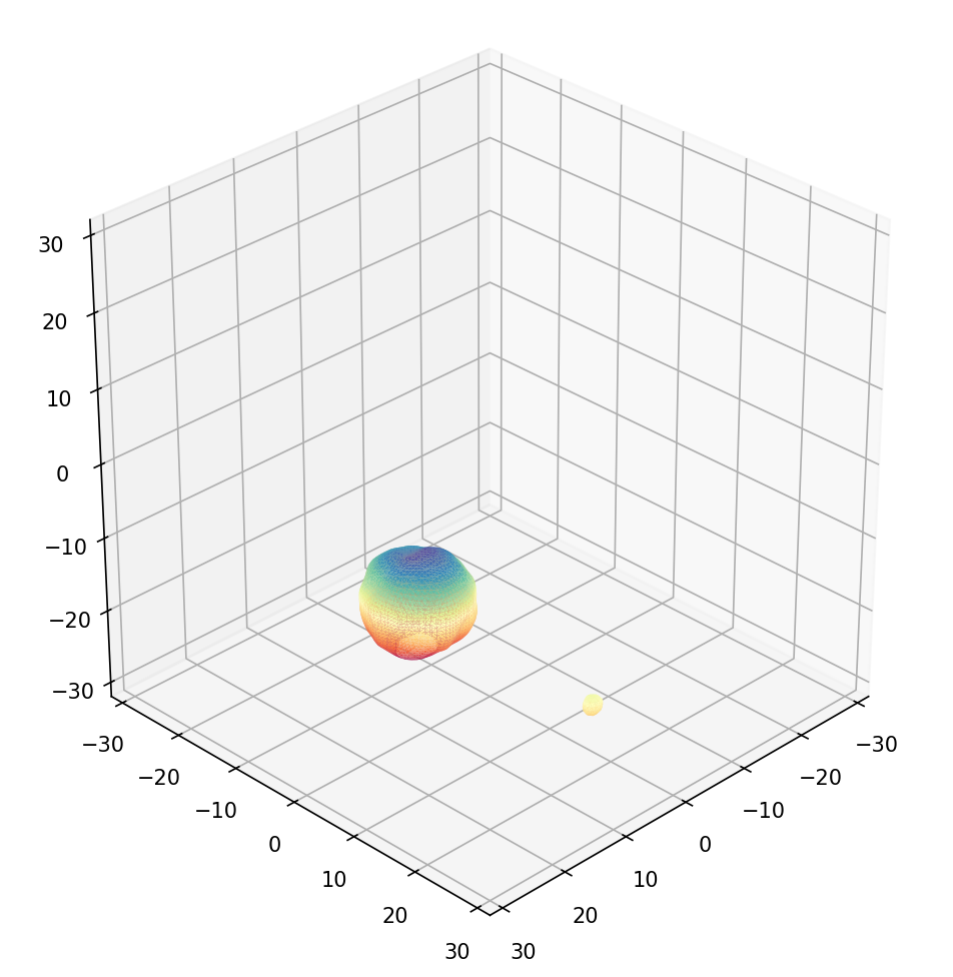}
        \caption{}
        \label{fig:3d_2800}
    \end{subfigure}

    \caption{Time evolution of total domain wall length (area) in 2D (a) and 3D (b). Panels (c)–(e) and (f)–(h) show snapshots of the field configuration at timeslices 40, 200, 280: (c,f) $t=40$, (d,g) $t=200$, (e,h) $t=280$. The departure from the early decay rate signals the transition to stable, high‑amplitude oscillon relics. Parameters are from ~\ref{app:2hdm} with mixing angle $\tan a=\tan b=0.85$.}
    \label{fig:domain_wall}
\end{figure*}

As shown in Fig.~\ref{fig:domain_wall}, after approximately $t=50.0$ in 2D and $t=250.0$ in 3D, the decay rate is significantly slower than the early stage prediction \cite{Battye_2020}. The system is set up with random initial conditions and periodic boundary conditions (toroidal space) \cite{BATTYE2025139311}, however, the periodic boundary condition gives that the radiating background particles (stochastic perturbations) may re-enter the system from the other side. As a result, the structure is immersed in a high-density collision condition, and the lifetime of the structure is highly reduced compared to what it could be in clean space. 

The 2D example shown in Fig.~\ref{fig:domain_wall} is used in the following simulations, as 3D analysis will take too long to run due to the complexity from the extra dimension in grid space. The result from Fig.~\ref{fig:domain_wall} shows that they have a similar pattern in producing long-lived bubbles so they should share the same mechanics.

\subsection{Forming Rate of Long-Lived Structures}

We would like to know if such an oscillon phenomenon exists in 2HDM only or can be reproduced in any simplified ${\mathbb{Z}_2}$-symmetry toy models for demonstrating the mechanism more clearly. We plot the terminal period of the decay curve in different models, in searching for the conditions required for such a structure to be produced.

In the introduction of 2-field and 4-field, as a comparison with 8-fields 2HDM in investigating the mechanics of the long-lived structure, and compare to complex doublet $\Psi_1$, $\Psi_2$ in 2HDM Lagrangian, we replace $\Psi_1$, $\Psi_2$ in equations (\ref{eq:lagrangian}) and (\ref{eq:potential}) with two real scalar fields or two complex scalar fields~\footnote{Both the two-complex-scalar and two-real-scalar models share the same $U (1)\times U (1)$ structure, so only one is discussed.} in representing 2-field or 4-field systems. In such cases, we only keep $\psi_1$ and $\psi_2$ (for the 2-field model), or $\psi_1$ through $\psi_4$ (for the 4-field model), and set all other fields in equations (\ref{eq:eom1}) and (\ref{eq:eom2}) to zero. 

For each field model (2-field, 4-field, 2HDM), we conducted extensive numerical surveys comprising hundreds of independent trials (system start with different random initial distributions). As we are only interested in the physical abnormal curve at the end of decay, we should only consider trials that end in single domain \cite{BATTYE2025139311} final states (all in one vacuum) only, since the periodic boundary condition makes the system into a torus. 

\begin{figure}   
    \centering
    \includegraphics[scale=0.5]{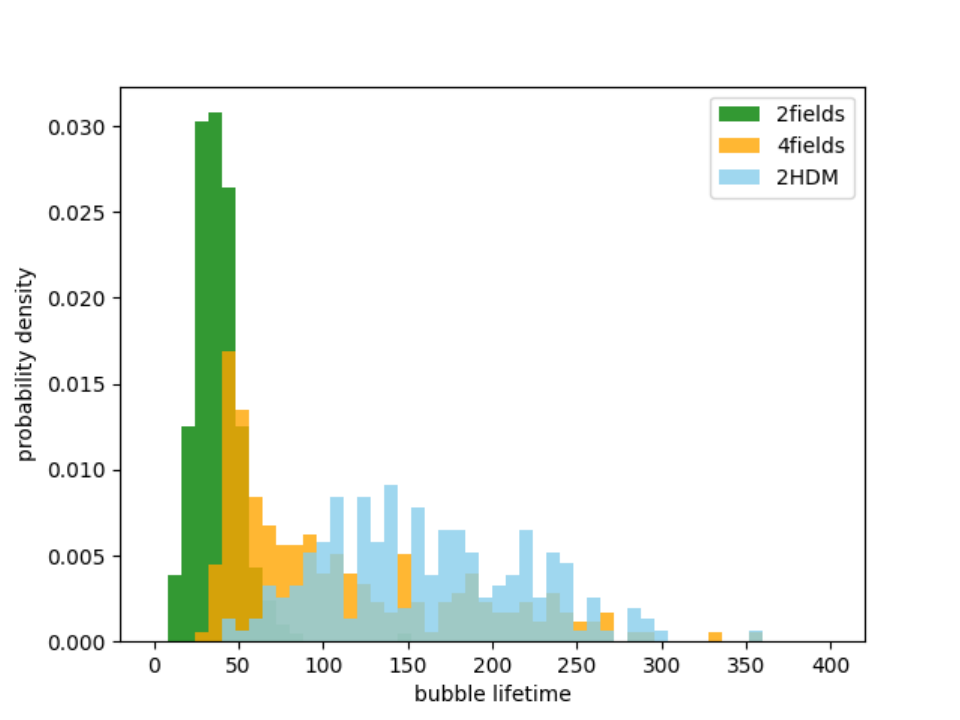}
    \caption{Histogram of bubble lifetimes in the 2‑field (two real scalars), 4‑field (two complex scalars) and full 2HDM (two complex scalar doublets). 300 independent trials were performed for each model; only trials that end in a single domain (around $60\%$ of all) are included. The 2‑field model serves as a minimal reference with no unitary symmetry and shows almost no long‑lived bubbles, confirming that at least two complex fields are required. Parameters are from ~\ref{app:2hdm} with mixing angle $\tan a=\tan b=0.85$.}
    \label{lifetime}
\end{figure}

Only when the total length of the domain wall network drops below the minimal length required for a stripe to wrap around the periodic box, the remaining walls necessarily form closed loops (bubbles) rather than extended stripes. This transition defines the onset of the bubble-dominated regime. We define 'bubble lifetime', which starts from when the domain wall length first falls below the minimum required to form a stripe in space (for a simulation of square with size $L$, the critical length is $L_{\mathrm{crit}} = 2L$), and ends at the wall length reaching zero for the first time. We neglect all trials with 'bubble lifetime' undefined to ensure that stripe modes are excluded. Therefore, in all three modes, the 'bubble lifetime' of each trial of simulation is plotted in a histogram of probability density against lifetime as shown in Fig.~\ref{lifetime}.

Previous research predicts that general systems are decaying at an increasing but certain rate \cite{BATTYE2025139311}, and the existence of bubbles can greatly slow down the decay to extend the lifetime of the structure, as shown in Fig.~\ref{fig:domain_wall}. Hence, the wider the spread of the histogram, the more bubbles are expected to exist, as the system is more divergent from the certain decay rate prediction. Therefore, Fig.~\ref{lifetime} shows many long-lived bubbles in 2HDM, fewer natural bubbles in the 4-field model, and no (or very rare) natural bubbles in the 2-field model.

The numerical results in Fig.~\ref{lifetime} indicate that the 4-field model, while producing fewer long-lived bubbles than the full 2HDM, still exhibits the phenomenon, while the two‑real‑scalar model serves as a minimal reference system with no unitary symmetry, highlighting that the emergence of long‑lived bubbles requires at least two complex fields, or unitary symmetry. This suggests that the 4-field model is the minimal framework capable of sustaining such oscillatory solutions. Its topological structure can be embedded into the 2HDM automatically, making it a useful toy model for illustrating the oscillon phenomenon in the full theory.


\section{Construction of the Ideal Long-Lived Structure}
\label{sec:ideal}
To isolate the essential mechanism of radiation suppression, we first consider a simplified toy model consisting of two complex scalar fields without gauge interactions. This model is not intended to represent the full 2HDM but serves as an analytically tractable system that captures the core nonlinear dynamics. Once the oscillon solution and its stability are understood in this reduced setting, we embed it back into the full 2HDM to confirm that the same configuration survives in the realistic theory.
\subsection{Oscillon Solution in Toy Model}
We first consider the minimal model that accommodates an oscillon, 4-real field system, expressed in terms of two complex scalars $\Psi_{1}$ and $\Psi_{2}$. In searching for a system that can maximize the structure lifetime, we identify a specific parameter subspace where high-order terms responsible for radiation are suppressed \cite{Amin_2010,Salmi:2012ta}. Under such conditions, parameters in the Lagrangian should additionally satisfy:

\begin{equation}
\begin{aligned}
    \mathcal{L} =& \left (\partial_\mu \Psi_1\right)^2   +\left (\partial_\mu \Psi_2\right)^2 -V (\Psi_1,\Psi_2)\\
    V (\Psi_1, \Psi_2) =& -{\mu^2} (\Psi_1^\dagger \Psi_1) - {\mu^2} (\Psi_2^\dagger \Psi_2)\\& + {\lambda} (\Psi_1^\dagger \Psi_1)^2 + {\lambda} (\Psi_2^\dagger \Psi_2)^2 \\&+ 2{\lambda} (\Psi_1^\dagger \Psi_1) (\Psi_2^\dagger \Psi_2)\\& + 2\lambda_4 \left(\mathrm{Re} (\Psi_1^\dagger \Psi_2)\right)^2, 
\end{aligned}
\end{equation}
Turning off $\lambda_4$, the potential is then enhanced to be $SU (2)$-invariant with $\Psi= (\Psi_1,\Psi_2)^T$ the doublet. But we need $\lambda_4\neq 0$ where reasons given in ~\ref{app:4field}. 

We may decompose the two complex fields $\Psi_1$ and $\Psi_2$ in terms of the real and imaginary parts, that $\Psi_1=\psi_1+i\psi_2$ and $\Psi_2=\psi_3+i\psi_4$, where the simplest solution that gives an oscillon is in the form of

\begin{equation}
\begin{aligned}
\psi_1=&+f_0\cos (\alpha)+f_1\cos (\omega t),\\
\psi_2=&-f_0\sin (\alpha)-f_1\sin (\omega t),\\
\psi_3=&-f_0\cos (\alpha)+f_1\cos (\omega t),\\ 
\psi_4=&+f_0\sin (\alpha)-f_1\sin (\omega t),
\label{eq:realfields}
\end{aligned}
\end{equation}

for four scalar fields $\psi_i$ and $\alpha$ is an arbitrary phase constant that reflects continuous vacuum degeneracy and does not affect the oscillon profile. And we expect when there are no higher-order terms ($\cos (2\omega t),\cos (3\omega t)$, etc.) up to infinity, there will be no radiation \cite{Amin_2010,Salmi:2012ta}.

The equation of motion can be rearranged by substituting terms in forms of $f_0$ and $f_1$ terms, hence we may obtain
 
\begin{equation}
\begin{aligned}
0 =& -\frac{d^2}{dt^2}\bigl(f_0\cos{\alpha} + f_1\cos(\omega t)\bigr) \\
&+ \nabla^2\bigl(f_0\cos{\alpha} + f_1\cos(\omega t)\bigr) + \mu\bigl(f_0\cos{\alpha} + f_1\cos(\omega t)\bigr) \\
&-2\lambda\bigl(f_0\cos{\alpha} + f_1\cos(\omega t)\bigr)\Bigl[\bigl(f_0\cos{\alpha} + f_1\cos(\omega t)\bigr)^2 \\
&\qquad\qquad + \bigl(-f_0\sin{\alpha} - f_1\sin(\omega t)\bigr)^2\Bigr] \\
&-2\lambda\bigl(f_0\cos{\alpha} + f_1\cos(\omega t)\bigr)\Bigl[\bigl(-f_0\cos{\alpha} + f_1\cos(\omega t)\bigr)^2 \\
&\qquad\qquad + \bigl(f_0\sin{\alpha} - f_1\sin(\omega t)\bigr)^2\Bigr] \\
&-2\lambda_4\bigl(-f_0\cos{\alpha} + f_1\cos(\omega t)\bigr)\bigl(f_0\cos{\alpha} + f_1\cos(\omega t)\bigr) \\
&\qquad \times \bigl(-f_0\cos{\alpha} + f_1\cos(\omega t)\bigr) \\
&-2\lambda_4\bigl(-f_0\cos{\alpha} + f_1\cos(\omega t)\bigr)\bigl(-f_0\sin{\alpha} - f_1\sin(\omega t)\bigr) \\
&\qquad \times \bigl(f_0\sin{\alpha} - f_1\sin(\omega t)\bigr),
\end{aligned}
\end{equation}
  
for the equation of motion of $\psi_1$ as an example.

Under this specific parameterization, the source terms for third and higher-order harmonics cancel exactly through trigonometric identities. Consequently, the equations of motion for the four-field system reduce to a closed set of coupled differential equations for the base frequency components $f_0$ and $f_1$.
\begin{equation}
    \begin{aligned}        
    0=&-\frac{d^2}{dt^2} (f_0\cos (\alpha)+f_1\cos (\omega t))\\&+\nabla^2 (f_0\cos (\alpha)+f_1\cos (\omega t))\\&+{\mu} (f_0\cos (\alpha)+f_1\cos (\omega t))\\&-2\lambda (f_0\cos (\alpha)+f_1\cos (\omega t)) (2f_0^2+2f_1^2)\\&-2\lambda_4 (-f_0\cos (\alpha)+f_1\cos (\omega t)) (-f_0^2+f_1^2),
    \end{aligned}
    \label{eq:cancellation}
\end{equation}
which can be decomposed based on the frequency spectrum ($\cos (\alpha)$ and $\cos (\omega t)$), the functions of $f_0$ and $f_1$ should satisfy
\begin{equation}
\begin{aligned}
        0=&\nabla^2f_0+{\mu^2}f_0- (4\lambda+2\lambda_4)f_0^3\\&- (4\lambda -2\lambda_4)f_0f_1^2,\\
        0=&\omega^2f_1+\nabla^2f_1+{\mu^2}f_1 - (4\lambda+2\lambda_4)f_1^3\\&- (4\lambda -2\lambda_4)f_1f_0^2.
        \label{eq:f0f1}
\end{aligned}
\end{equation}
and the equations of motion for the other three fields yield identical results.

\begin{figure*}   
    \centering
    \includegraphics[scale=0.4]{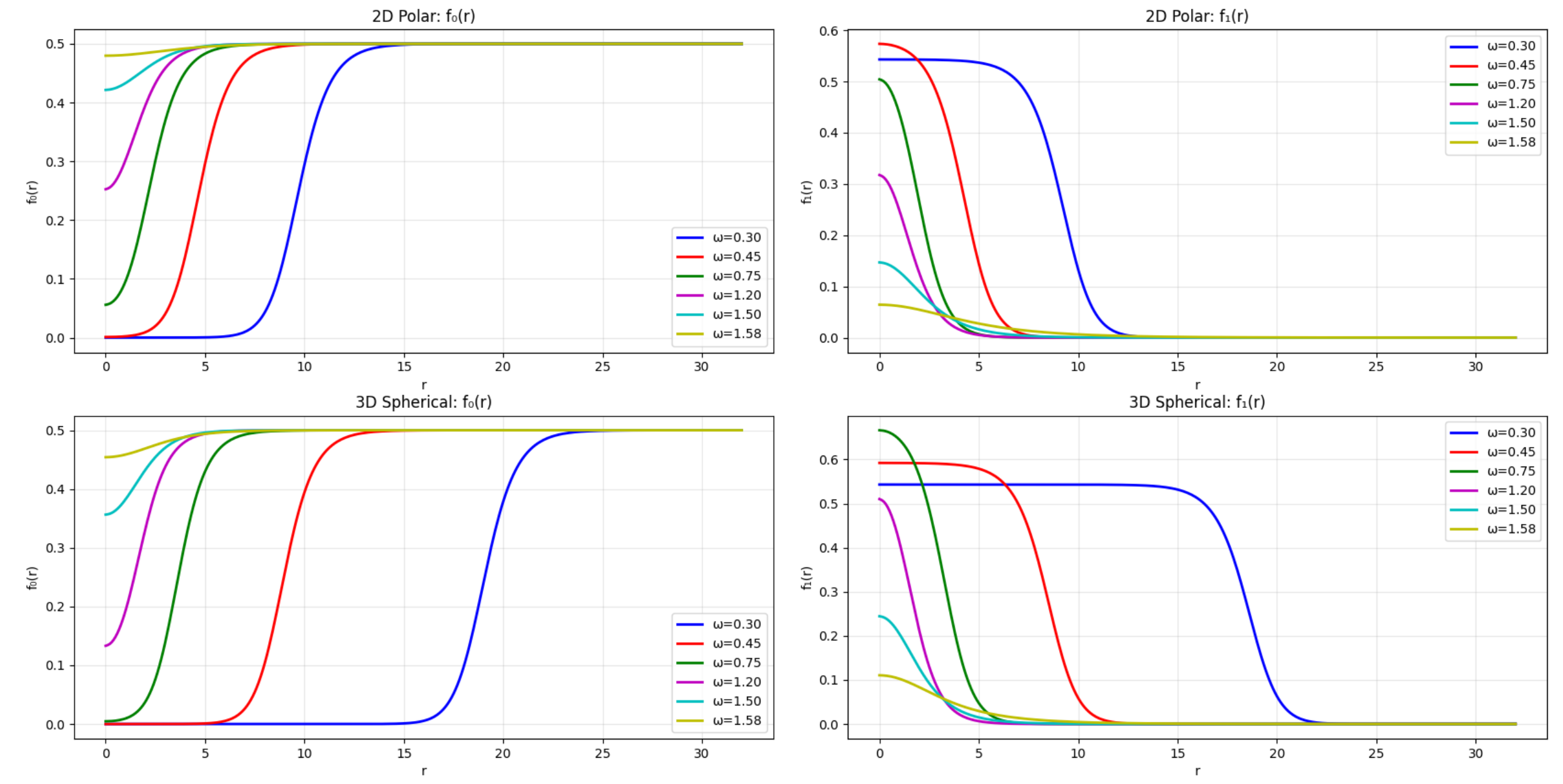}
    \caption{Plot of $f_0$ and $f_1$ in 2D and 3D space, with choice of parameter $\mu= 0.5$, $\lambda= 1.78$, and $\lambda_4 = -2.56$ as in ~\ref{app:2hdm} with full mixing. The radius where the $f_0$ and $f_1$ curves cross (gives $R^1=0$) corresponds to the domain wall location. There is a maximum frequency $\omega_m=1.6$ where any system should have a frequency below that $\omega<\omega_m$. Each frequency selects a different radius, demonstrating that the oscillon exists as a continuous family parameterized by $\omega$, with no preferred or characteristic size; all frequencies in this family are equally stable. See ~\ref{app:git} for the code used to find these solutions.}
    \label{f0f1}
\end{figure*}

At far infinity ($r\to\infty$), we impose the boundary conditions $f_1=0$, $f_0=\sqrt{\mu^2/(4\lambda+2\lambda_4)}$, and $\nabla f_0=\nabla f_1=0$. After applying those boundary conditions, the shape of the functions is plotted in Fig.~\ref{f0f1}.

We numerically find the oscillon configurations and demonstrate them in Fig.~\ref{f0f1}. The profiles depend continuously on $\omega$, with each frequency selecting a different radius. There is no preferred or characteristic radius; the entire family $\omega \in (0, \omega_m)$ is linearly stable. Consequently, all frequencies share the same parametrically infinite lifetime in the ideal radiation-suppressed limit. The behavior of the profiles as $\omega$ varies is as follows:
\begin{itemize}
    \item The maximum frequency of the system, $\omega_m$, also satisfies the relation, namely that $M_h^2=V'' (\textbf{vac})=\mu^2 \frac{2|\lambda_4| }{2\lambda+\lambda_4}=2.56=\omega_m ^2$, where $M_h$ is the mass term of the system given in ~\ref{app:2hdm}. As $\omega$ approaches $\omega_m$, the functions have smaller amplitude with a wider peak, approaching horizontal lines (vacuum).
    \item The profiles described above form a continuous one-parameter family. As $\omega$ varies, the shape changes continuously, but all members are equally stable. One type is domain-wall-like for the low $\omega$: the center of the oscillon is near one vacuum with $f_0 (0)$ almost vanishes while $f_1 (0)$ approaches $v$; $f_0 (r)$ and $f_1 (r)$ grow and decrease towards the outer region, respectively, interpolating in another vacuum. One may notice that, for small $\omega$, there is an obvious flat-top structure (with $\nabla ^2f_1 (r)\approx 0$) with certain width within which $f_0 (r)\to 0$ and the oscillating  amplitude keeps constant $f_1^2 (r)\approx \frac { \mu^2-  \omega^2} {2 (2\lambda+\lambda_4)}$. Thus, the width narrows as the frequency increases; at the same time, the oscillating amplitude $f_1 (r)$ increases.   
    \item As we increase $\omega$ toward $\omega_m$, the width of the Gaussian peak extends, but the height (oscillating amplitude) shrinks. The corresponding solutions are not domain-wall-like, because the fields are just oscillating slightly around one vacuum, which is called a small-amplitude oscillation.
    \item As $\omega\rightarrow 0$, the inner region of the opposite vacuum goes infinitely large, and that space fully goes into the opposite vacuum. This shows that the large-amplitude oscillon can continuously transition into a wall structure, that the nature of this structure is indeed a closed wall separating two vacua shown in Fig.~\ref{fig:domain_wall}, and a large-scale bubble-like closed domain wall structure could persist on cosmological timescales.
\end{itemize}

\begin{figure*} 
    \centering
    {
    {\includegraphics[trim={0cm 0cm 0cm 0.3cm}, clip, scale=0.4]{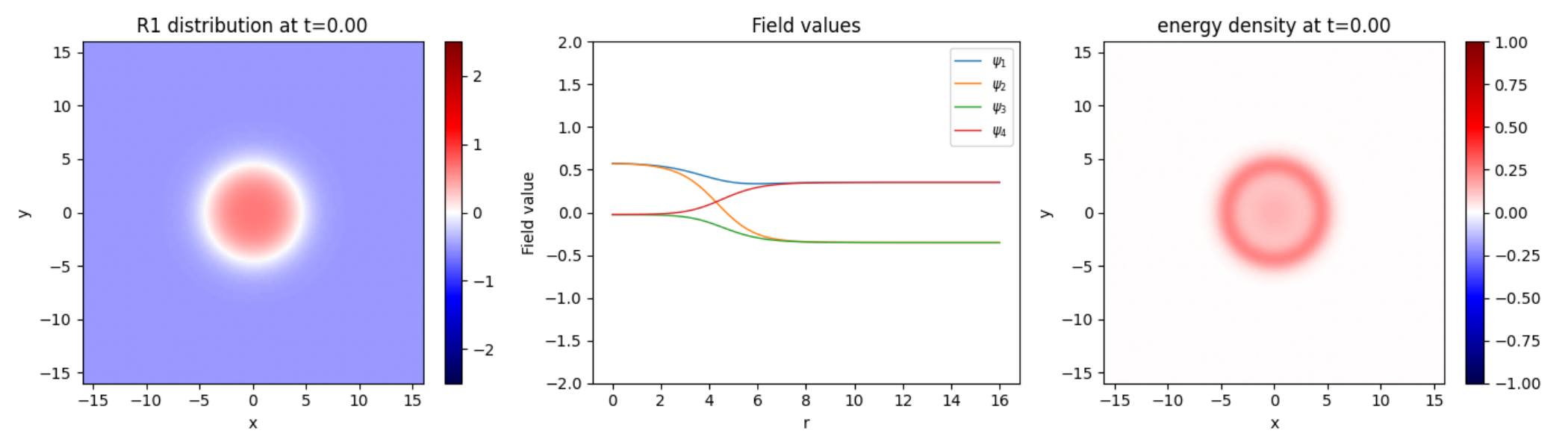}}
    {\includegraphics[trim={0cm 0cm 0cm 0.3cm}, clip, scale=0.4]{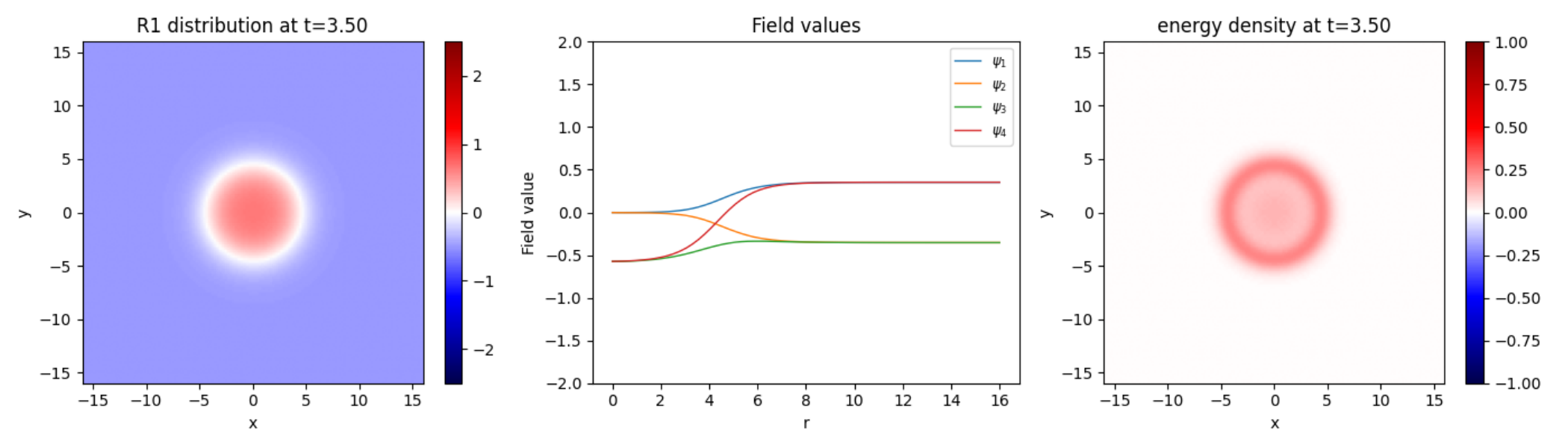}}
    {\includegraphics[trim={0cm 0cm 0cm 0.3cm}, clip, scale=0.4]{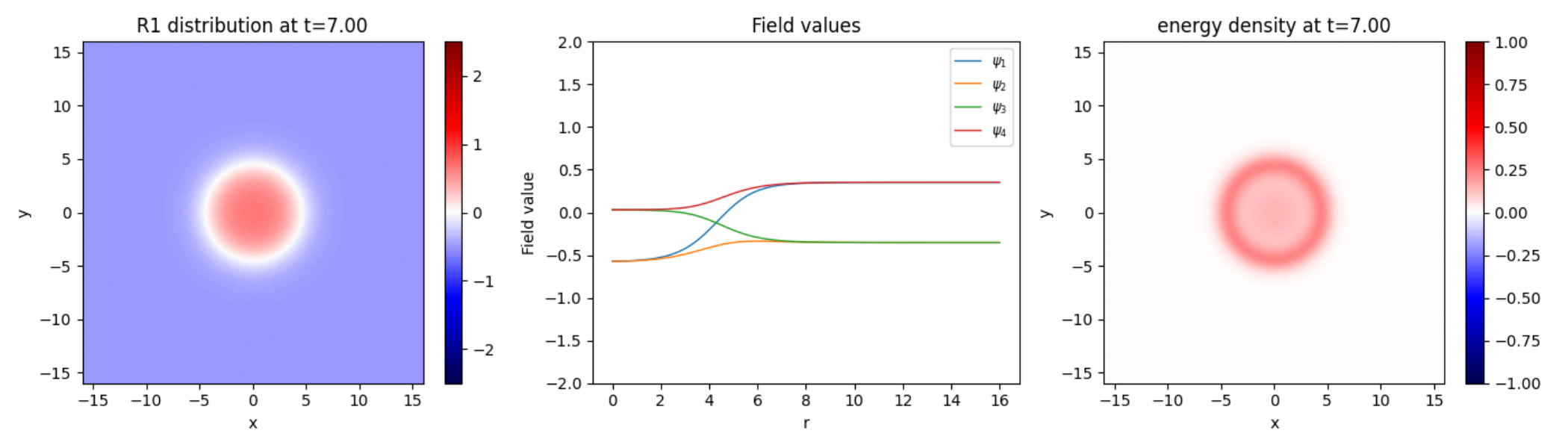}}
    {\includegraphics[trim={0cm 0cm 0cm 0.3cm}, clip, scale=0.4]{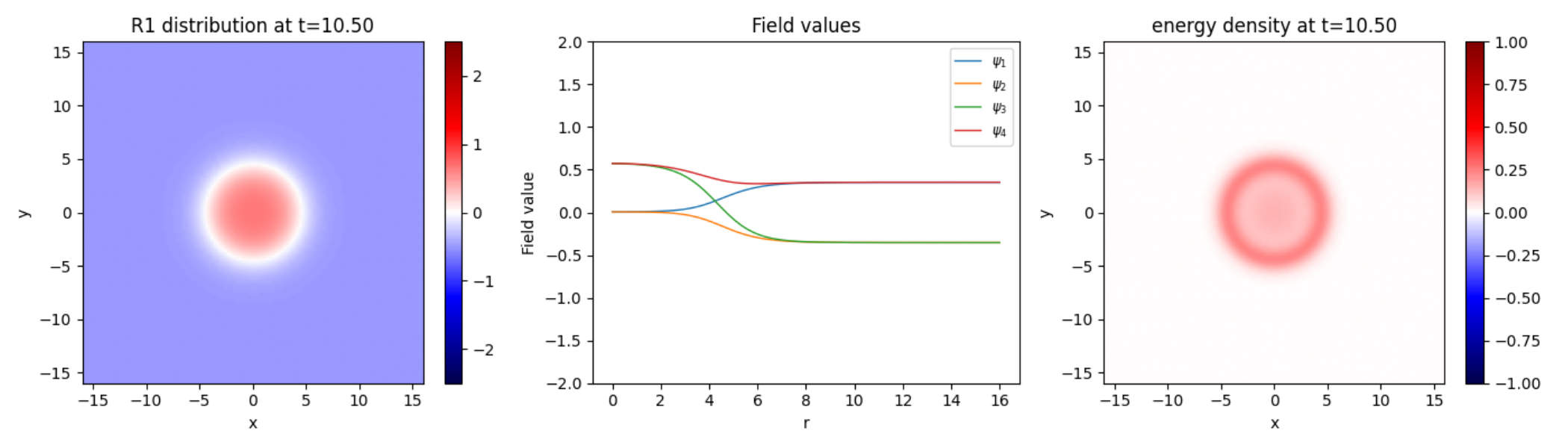}}
    }

    \caption{The plot of time slice at rotating phase of fields in $0$, $\pi/2$, $\pi$, $3\pi/2$. with the $R^{1}$ distribution (left column),  radial field distribution (middle column), and energy distribution (right column) plot. The plots use a constant phase $\alpha=\pi/4$, and other parameters used are from ~\ref{app:2hdm} with full mixing (mixing angle $a=b=\pi/4$). The sponge-layer absorbing boundary condition is used to replace periodic ones. Full evolving region has radius R=32 in this simulation, though only the center part is displayed. Time-evolution animations and the associated numerical codes are publicly available in the GitHub repository (see ~\ref{app:git}).}
    \label{in_bubble}
\end{figure*}

After successfully obtaining the mathematical expression of the structure, we reproduce it in the lattice grid simulation. Unlike the periodic boundary conditions used for the early-universe simulations with dense background noise, we now employ the sponge-layer absorbing boundary condition \cite{sponge} to eliminate incoming radiation and ensure that the structure evolves in a clean environment. As shown in Fig.~\ref{in_bubble}, fields involved in the simulation are made in a purely circular symmetrical pattern, which means that bubbles can exist without any circular motion. We found the field distribution in the simulation shown in Fig.~\ref{in_bubble} is exactly the same as predicted by the previous analysis.

An alternative perspective on this configuration emerges when we rewrite the two complex fields in a form that reveals a Q‑ball interpretation \cite{Coleman:1985ki,azatov2024qballperturbationsdetailslinear}. Noting that the ansatz (equation (\ref{eq:realfields})) can be compactly expressed as
\begin{equation}
\begin{aligned}
\Psi_1 =& f_0 (r) e^{-i\alpha} + f_1 (r) e^{-i\omega t},\\
\Psi_2 =& f_0 (r) e^{-i\alpha} + f_1 (r) e^{i (\pi-\omega t)},
\end{aligned}
\end{equation}
where the relative phase $e^{i\pi}$ accounts for the sign difference in the oscillating parts. Introducing the linear combinations $\Psi_\pm = (\Psi_1 \pm \Psi_2)/\sqrt{2}$ simplifies the system to
\begin{equation}
\Psi_+ = \sqrt{2}\, f_0 (r) e^{-i\alpha},\qquad 
\Psi_- = \sqrt{2}\, f_1 (r) e^{-i\omega t}.
\end{equation}
Thus $\Psi_+$ is static and $\Psi_-$ rotates uniformly in the complex plane, with a conserved global $U (1)$ charge whose density is given by
\begin{equation}
\rho (r) = i \sum_{a=1}^{2} \left( \Psi_{a}^{\dagger} \partial_{t} \Psi_{a} - \left( \partial_{t} \Psi_{a}^{\dagger} \right) \Psi_{a} \right)= 4\omega\, f_1 (r)^2.
\end{equation}

The charge density $\rho(r) = 4\omega f_1(r)^2$ arises from a conserved global $U(1)$ Noether current associated with the phase rotation symmetry of the complex scalar fields. The charge is localized where $f_1(r)$ is nonzero, i.e., throughout the bubble interior, with total charge $Q=4\omega \int f_1^2 \, dV$. This confirms the Q-ball (or adjusted FLS structure \cite{azatov2024qballperturbationsdetailslinear}) nature of the oscillon, and explains its longevity.

This Q-ball interpretation is closely related to the charge-breaking vacuum condensates ($v_+ \neq 0$) found in static domain wall solutions of the 2HDM \cite{Battye:2025qbb,Sassi:2024dw}. In those works, the non-trivial topology of the vacuum manifold allows charge accumulation inside the wall. Our oscillons are time-dependent, but the underlying mechanism is the same: the presence of a conserved $U(1)$ current permits localized charge. The key difference is the spatial distribution: in static domain walls, the condensate is localized on the wall surface; in our oscillon, as shown by the $f_1(r)$ curves in Fig.~\ref{f0f1}, the condensate is distributed throughout the bubble interior. The charge is carried by an oscillating configuration rather than a static condensate, and stabilized by the frequency $\omega$ as a free parameter rather than by a static topological constraint.

\subsection{Stability of the Structure}
To determine whether the oscillon is stable against small perturbations, we perform a Floquet analysis, which examines the growth of perturbations over one complete oscillation period. An oscillon solution requires that the perturbation should not expand over time.

\begin{figure}   
    \centering
    \includegraphics[scale=0.275]{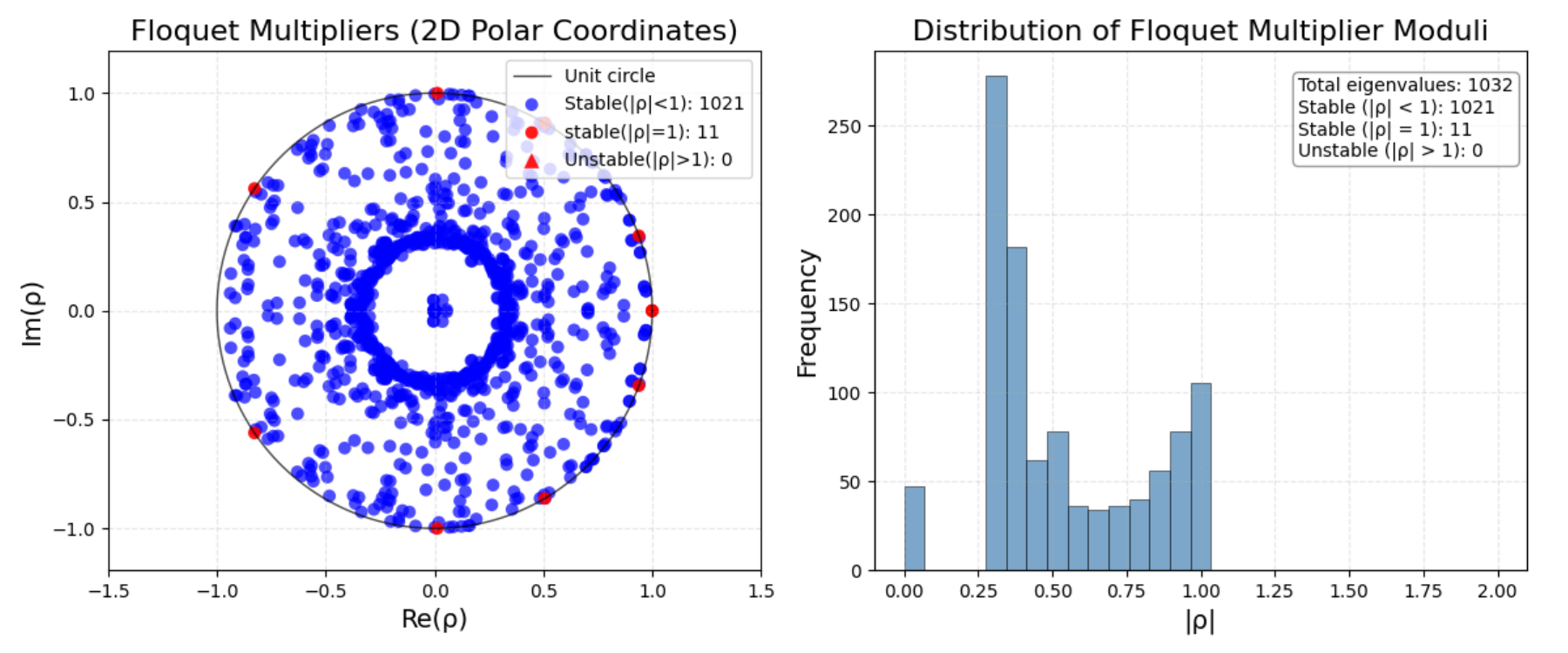}
    \includegraphics[scale=0.275]{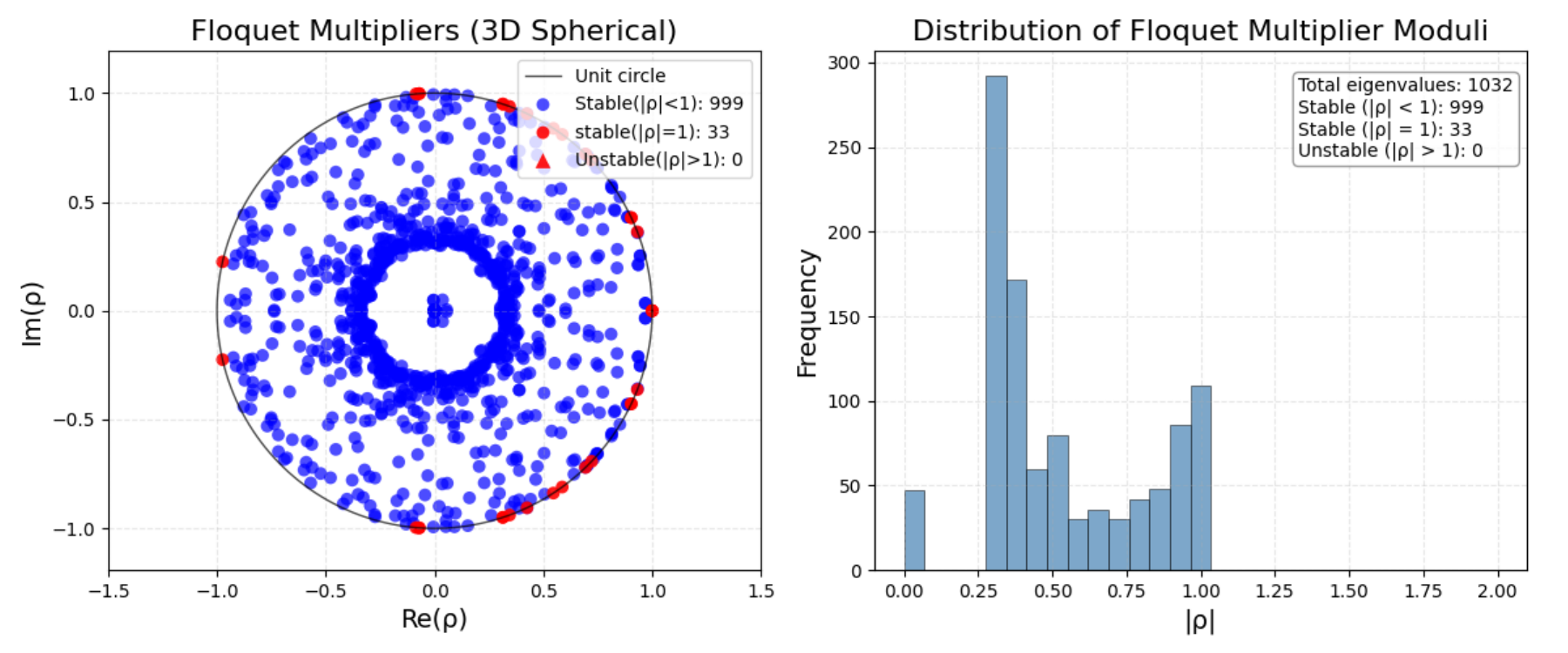}
    \caption{Plot of all eigenvalues (Floquet multipliers) from the matrix of Floquet analysis on the complex plane (left), and histogram of moduli (right), for 2D (up) and 3D (down) systems. Points labeled as 'equal to 1' include values within $1\pm10^{-5}$ due to the consideration of decimal uncertainties from grids and iteration. Parameters and background are chosen from Fig.~\ref{f0f1} with $\omega=0.45$, as a representative intermediate frequency where the oscillon exhibits a clear domain-wall-like structure, though analysis gives the same stable statement under other parameter choice. See ~\ref{app:git} for analysis codes.}
    \label{floquet}
\end{figure}

For the mechanics and system setup of Floquet analysis discussed in ~\ref{app:floquet}, and for $\omega=0.45$ curve in Fig.~\ref{f0f1} as an example~\footnote{We have also tried to analyze the stability from some other setup mode, like all other plots with different frequencies in Fig.~\ref{f0f1}, their analysis results are all stable and see ~\ref{app:git} for codes.}, Floquet analysis in circular symmetry perturbation modes gives the result: some mode gives eigenvalues smaller than or equal to (with less than $\pm10^{-5}$ uncertainty) 1, as shown in Fig.~\ref{floquet}, indicating the system will become stable against first-order small-scale perturbations over time.

Floquet analysis in Fig.~\ref{floquet} indicates that there are no modes that expand over time in radial perturbation; the system is stable in radial modes. This stability holds uniformly for all frequencies $\omega \in (0, \omega_m)$. Under non-radial perturbation modes, from which perturbation is decomposed into a summation of modes with different angular spectrum $\delta (t,r,\theta) = \sum_l \delta_l (t,r) e^{i l \theta}$, where each term in them gives a shift to the potential by 
\begin{equation}
\begin{aligned}
    V_{eff} (r,l) = V (r) + l^2/r^2 \quad\text{ (2D),} \\V_{eff} (r,l) = V (r) + l (l+1)/r^2  \quad\text{ (3D).}
\end{aligned}
\end{equation}

For all $\partial^2\delta/\partial t^2 = \left(\nabla^2 - V_{eff} (r,l)\right) \delta$ where there is $V_{eff} (r,l) > V (r)$ at $l\geq1$, these non-radial modes are energetically more confined than the radial ($l=0$) mode. Since the radial modes are already proven stable (as shown in Fig.~\ref{floquet}), the system is consequently stable against perturbations in any angular momentum sectors.

\subsection{Embedding the Oscillon into the Full 2HDM}

While the 4-field subsystem provides the minimal framework for non-radiating oscillons, it is imperative to verify if these configurations remain viable within the full structure of the 2HDM. We therefore map the previous ansatz onto the eight-field field space of the full complex doublet potential, to ensure the cancellation of high-frequency terms.

To satisfy the cancellation condition of radiation terms, we use a set of parameter values from ~\ref{app:2hdm} \cite{Battye_2020,Battye_2011,Hespel_2014}, but modifying the mixing parameter to $a=b=\pi/4$, to have $\mu^2=0.5$, $\lambda=1.78$, and $\lambda_4=-2.56$. Then, parameters in the Lagrangian can be rearranged into \cite{Battye_2020,BATTYE2025139311}
\begin{equation}
\begin{aligned}
    V (\Psi_1, \Psi_2) =& -{\mu^2} (\Psi_1^\dagger \Psi_1) - {\mu^2} (\Psi_2^\dagger \Psi_2) \\&+{\lambda} (\Psi_1^\dagger \Psi_1)^2 + {\lambda} (\Psi_2^\dagger \Psi_2)^2 \\& + {2\lambda} (\Psi_1^\dagger \Psi_1) (\Psi_2^\dagger \Psi_2)\\& + {2\lambda_4} \left(\mathrm{Re} (\Psi_1^\dagger \Psi_2)\right)^2 + 0 \left(\mathrm{Im} (\Psi_1^\dagger \Psi_2)\right)^2.
\end{aligned}
\end{equation}
The higher frequency terms $\cos^2 (\omega t)$ and $\sin^2 (\omega t)$ will vanish in the same way as in the 4-field system discussed in the Section 4, and those full-mixing parameter values will be used in the rest of this research.

Like the solution in the 4-field system, the two complex fields doublet of the 2HDM,
\begin{equation}
    \Psi_1=\begin{pmatrix}\psi_1+i\psi_{3}\\ \psi_5+i\psi_{7}\end{pmatrix},\quad\quad\Psi_{2}=\begin{pmatrix}\psi_{2}+i\psi_4\\ \psi_{6}+i\psi_8\end{pmatrix},
\end{equation}
can be expanded in the form of $f_0$ and $f_1$ to fit the solution of oscillon
\begin{equation}
\begin{aligned}
    \psi_1=&+\left (f_0\cos (\alpha_1)+f_1\cos (\omega t)\right)\cos (\beta),\\
    \psi_2=&-\left (f_0\cos (\alpha_1)+f_1\cos (\omega t)\right)\cos (\beta),\\
    \psi_3=&-\left (f_0\sin (\alpha_1)-f_1\sin (\omega t)\right)\cos (\beta),\\
    \psi_4=&+\left (f_0\sin (\alpha_1)-f_1\sin (\omega t)\right)\cos (\beta),\\
    \psi_5=&+\left (f_0\cos (\alpha_1)-f_1\cos (\omega t+\theta_\alpha)\right)\sin (\beta),\\
    \psi_6=&-\left (f_0\cos (\alpha_1)-f_1\cos (\omega t+\theta_\alpha)\right)\sin (\beta),\\
    \psi_7=&+\left (f_0\sin (\alpha_1)+f_1\sin (\omega t+\theta_\alpha)\right)\sin (\beta),\\
    \psi_8=&-\left (f_0\sin (\alpha_1)+f_1\sin (\omega t+\theta_\alpha)\right)\sin (\beta),
\end{aligned}
\end{equation}
where $\alpha_1$, $\alpha_2$, $\beta$ and $\theta_\alpha$ are constant and arbitrary phase (though those phases are not important due to gauge symmetry), $\beta=0$ leads to previous 4-field system. Under such a setup, take those for $\psi_1$ as an example among eight. The equation of motion of $\psi_1$ can then be rearranged by substituting by $f_0$ and $f_1$ as
\begin{equation}
    \begin{aligned}        
    0=&-\frac{d^2}{dt^2} (f_0\cos (\alpha)+f_1\cos (\omega t))\cos (\beta)\\&+\nabla^2 (f_0\cos (\alpha)+f_1\cos (\omega t))\cos (\beta)\\&+{\mu} (f_0\cos (\alpha)+f_1\cos (\omega t))\cos (\beta)\\&-2\lambda (f_0\cos (\alpha)+f_1\cos (\omega t)) (2f_0^2+2f_1^2)\cos (\beta)\\&-4\lambda_4 (-f_0\cos (\alpha)+f_1\cos (\omega t)) (-f_0^2+f_1^2)\cos (\beta),
    \end{aligned}
\end{equation}
where all other phase parameters are canceled in trigonometric. Then we can repeat separating terms based on frequency spectrum step as we did in the 4-field system, and we will end up with a set of equations to the shape of $f_0$ and $f_1$,

\begin{equation}
    \begin{aligned}  
        0=&\nabla^2f_0+{\mu}f_0-2\lambda f_0 (2f_0^2+2f_1^2)\\&+2\lambda_4f_0 (-f_0^2+f_1^2),\\
        0=&\omega^2f_1+\nabla^2f_1+{\mu}f_1-2\lambda f_1 (2f_0^2+2f_1^2)\\&-2\lambda_4f_1 (-f_0^2+f_1^2),
    \end{aligned}
\end{equation}
which is exactly the same as equation (\ref{eq:f0f1}) and we can keep using values plotted in Fig.~\ref{f0f1}. The derivation starting from any other seven fields gives the same results. Like the theory of the 4-field system, the system is valid without any radiation terms; that is, the solution is also radiation‑suppressed.~\footnote{GIF file plotted for 2HDM field motion has been uploaded into GitHub (see ~\ref{app:git}).}

\begin{figure}   
    \centering
    \includegraphics[scale=0.275]{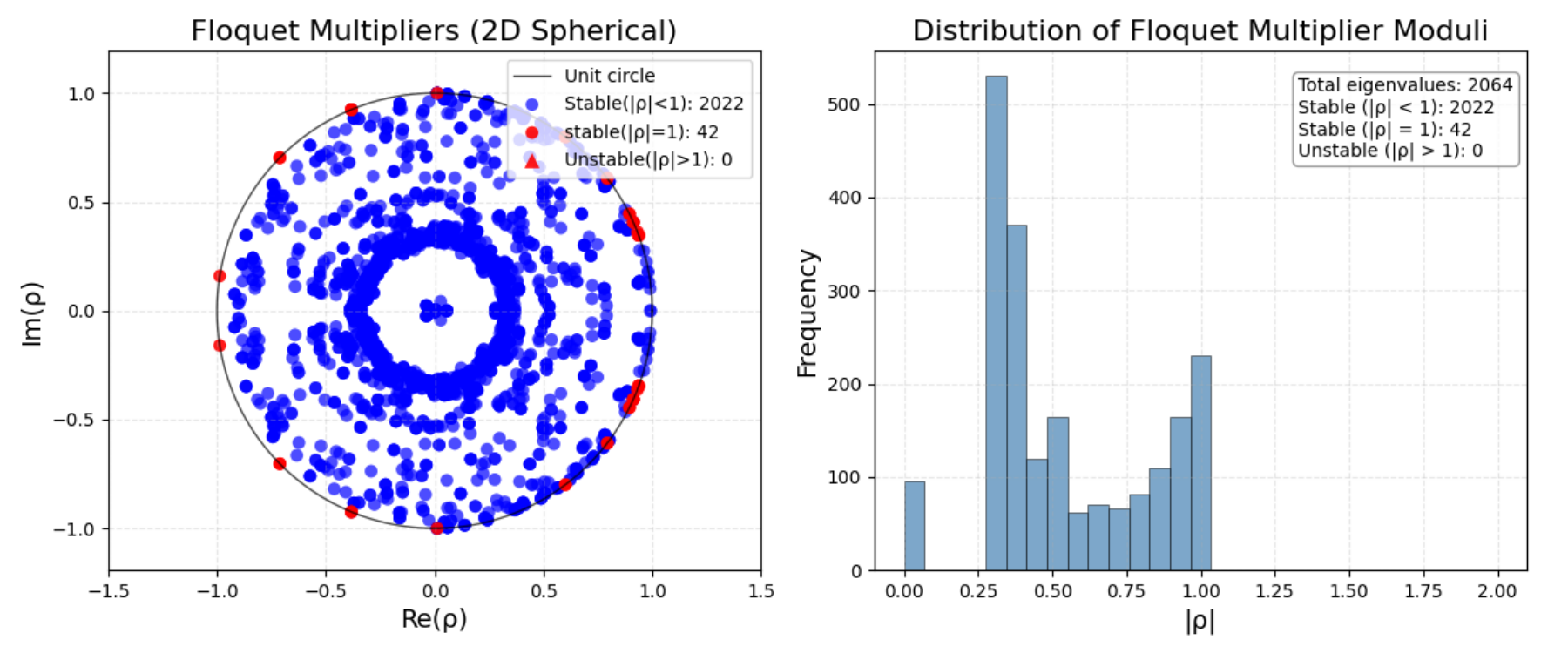}
    \includegraphics[scale=0.275]{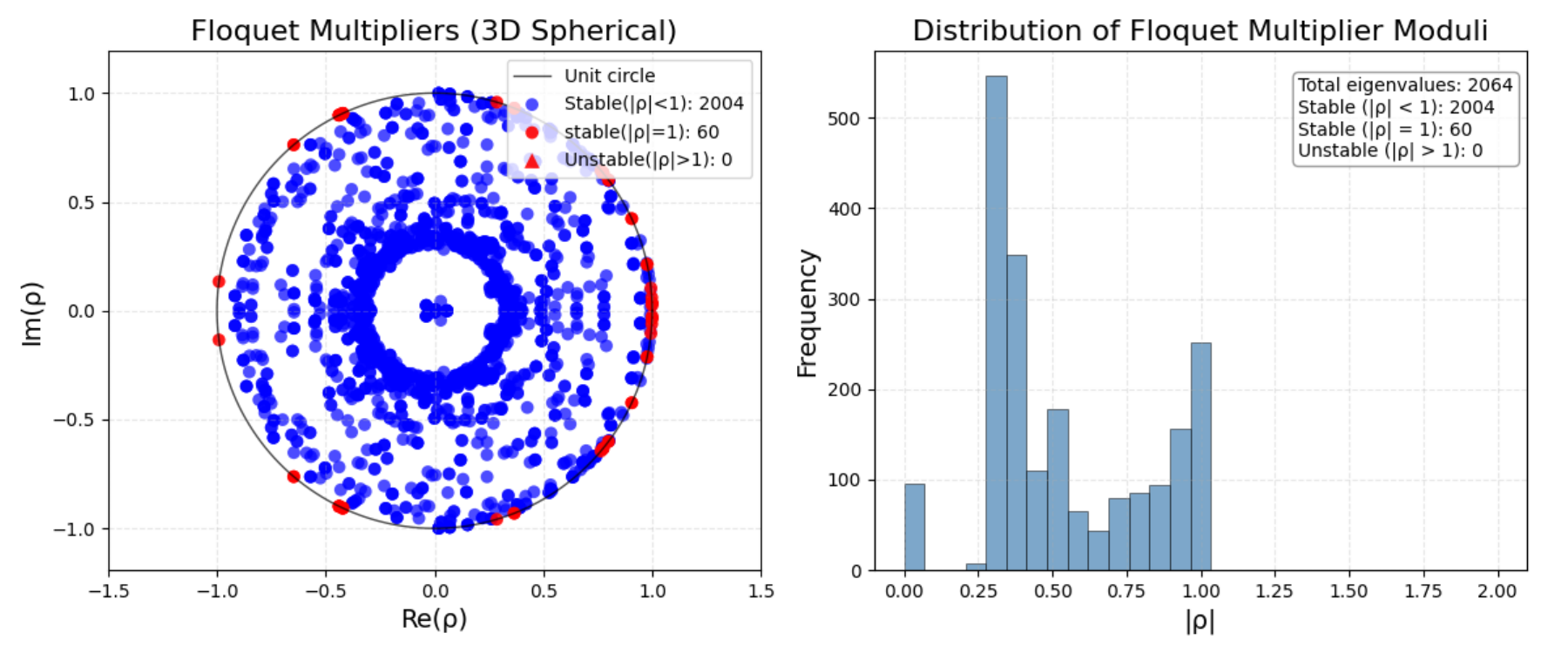}
    \caption{Floquet multipliers for the full 2HDM, with background solution taken at $\omega=0.45$ as an example, and $\alpha_1=\alpha_2=\beta=\pi/4$, and $\theta_\alpha=0$. Other parameters are from ~\ref{app:2hdm} with full mixing (mixing angle $a=b=\pi/4$). Like in the analysis in the 4-field system, further plots show the system is stable for all profiles shown in Fig.~\ref{f0f1}, or $\alpha_1$, $\alpha_2$, or $\beta$ chosen.}
    \label{floquet-2hdm}
\end{figure}

After finishing the 2HDM setup, we repeat the Floquet analysis of stability to compare it with the 4-field system. As shown in Fig.~\ref{floquet-2hdm}, like in the 4-field system, Floquet analysis also confirms the linear stability.

\section{Conclusions}


Though the current simulations are performed in flat spacetime to isolate the intrinsic dynamics of radiation suppression, the remarkable longevity and Q-ball-like nature of these oscillons suggest significant robustness. From the unexpected decay curve of the system starting with random initial conditions in previous research, we observe an oscillatory solution that exists as a long-lived, bubble-like closed domain wall, and such oscillon structure has been reproduced successfully in the artificial setup in 4-field theory as well as in 2HDM, with oscillating in one vacuum manifold inside and stationary at another vacuum outside. The charge-carrying nature of these oscillons also connects them to the charge-breaking condensates recently classified in static 2HDM domain walls \cite{Battye:2025qbb,Sassi:2024dw}, suggesting a unified picture of localized charge accumulation in the 2HDM vacuum manifold. Thus, the charge-breaking condensate is not only present in static walls but also in our time-dependent oscillon solutions, manifesting as the conserved $U(1)$ charge carried by the oscillating field configuration. Further investigation of 2HDM oscillon evolution within a full FRW background is reserved for future work. 

We also found that, when the full-mixing condition $\tan a=\tan b=1$ is satisfied, the structure becomes highly radiation-suppressed. This condition defines a continuous locus in the 2HDM parameter space (with masses and $\tan\beta$ free), and the lifetime can be made parametrically long. In the ideal limit, it is expected to exceed cosmological timescales. The computer modeling results are consistent with our predictions. However, while the mathematical stability is demonstrated, the precise topological protection mechanism warrants further investigation, which will be the subject of future work. Besides, even if these parameter choices are found not favored by collider constraints, the mathematical structure could be realized in condensed matter systems, where interaction parameters could be manually adjusted to fit the requirement.

There are also some limitations of using classical wave motion to simulate quantum field theories. Although the mechanics share some similarities, there are still differences between them. Our classical field simulation can be used as a reference, but one should bear in mind that it might be different from realistic physical behaviors. Long‑lived oscillons that survive until today could contribute to dark matter. Their formation and decay would also produce a stochastic gravitational wave background, potentially detectable by future observatories such as LISA or the Einstein Telescope. These possibilities warrant further investigation but lie beyond the scope of the present paper. Future work should explore the quantum stability of these oscillons, their potential role in dark matter production, and the gravitational wave signatures from their formation and decay in the early universe. Additionally, analogous structures may be realizable in cold-atom systems, providing a laboratory for studying non-equilibrium field theory.

\begin{acknowledgements}
I would like to express my thanks to Prof. Zhaofeng Kang for his enthusiastic support and invaluable guidance throughout this research, and to Prof. Kecheng Wang and Prof. Yinnan Mao for their help in improving my knowledge in cosmology. I am also grateful to Prof. Richard A. Battye and Dr. Steven J. Cotterill for their guidance and supervision during my master's research project, which was the former version of this research. 

Artificial intelligence has been used to assist in coding and improve grammar.
\end{acknowledgements}

\appendix

\section{Parameters and masses in the four-field system}
\label{app:4field}

After reducing four degrees of freedom, the four‑field system predicts fewer particles than the 2HDM. The Lagrangian density can be written in terms of tensors:
\begin{equation}
\mathcal{L} = (\partial_{\mu}\Psi_{i})(\partial^{\mu}\Psi_{i}) - V(\underline{\Psi}),
\end{equation}
where the array $\underline{\Psi}$ decomposes into its elements $\Psi_{i}$, and the Lagrangian is invariant under the symmetry group transformation
\begin{equation}
\Psi_{i} \rightarrow \Psi_{i}' = \Psi_{i} + i\theta^{a}T_{ij}^{a}\Psi_{j},
\end{equation}
for a Lie group generator $T^{a}$ with matrix representation $T_{ij}^{a}$, of the Lie group $G=SO(n)$ of $n$ dimensions (number of fields), with infinitesimal change $\theta^{a}$. Since the potential $V(\underline{\Psi})$ is invariant under this transformation, $\delta V = V(\Psi)-V(\Psi')=0$, so $\delta V = i\theta^{a}T_{ij}^{a}\Psi_{j}=0$, which yields $T_{ij}^{a}\Psi_{j}=0$.

Taylor expanding $\underline{\Psi}$ around a vacuum solution $\underline{v}$ by $\underline{\Psi}=\underline{v}+\underline{\psi}$, the first order term vanishes because $T_{ij}^{a}\Psi_{j}=0$, so the expansion is
\begin{equation}
V(\underline{\Psi}) = V(\underline{v}) + \frac{1}{2}M_{ij}^{2}\psi_{i}\psi_{j} + \mathcal{O},
\end{equation}
where
\begin{equation}
M_{ij}^{2} = \frac{\partial^{2}V}{\partial\Psi_{i}\partial\Psi_{j}}\bigg|_{\underline{\Psi}=\underline{v}},
\end{equation}
and $M_{ij}^{2}$ is the mass matrix of the fields $\psi_{i}$.

Choosing the field basis
\begin{equation}
\Psi_{1} = v+\psi_{1}+ia_{1},\qquad \Psi_{2} = v+\psi_{2}+ia_{2},
\end{equation}
with scalar fields $\psi_{1},\psi_{2},a_{1},a_{2}$ and the vacuum $v=f_{0}=\sqrt{\frac{\mu^{2}}{4\lambda+2\lambda_{4}}}$, the total potential expands to
\begin{equation}
\begin{aligned}
V(\Psi_{1},\Psi_{2}) = & -\mu^{2}\bigl((v+\psi_{1})^{2}+a_{1}^{2}\bigr) - \mu^{2}\bigl((v+\psi_{2})^{2}+a_{2}^{2}\bigr) \\
& +\lambda\bigl((v+\psi_{1})^{2}+a_{1}^{2}\bigr)^{2} + \lambda\bigl((v+\psi_{2})^{2}+a_{2}^{2}\bigr)^{2} \\
& +2\lambda\bigl((v+\psi_{1})^{2}+a_{1}^{2}\bigr)\bigl((v+\psi_{2})^{2}+a_{2}^{2}\bigr) \\
& +2\lambda_{4}\bigl[(v+\psi_{1})(v+\psi_{1})+a_{1}a_{2}\bigr]^{2}.
\end{aligned}
\end{equation}
Extracting the corresponding parts of the matrix and separating them gives
\begin{equation}
M_{h,H}^{2} = \frac{1}{2}\frac{\partial^{2}V}{\partial\psi_{i}\partial\psi_{j}} = \begin{pmatrix}
4\lambda v^{2} & 4(\lambda+\lambda_{4})v^{2}\\
4(\lambda+\lambda_{4})v^{2} & 4\lambda v^{2}
\end{pmatrix},
\end{equation}
\begin{equation}
M_{A}^{2} = \frac{1}{2}\frac{\partial^{2}V}{\partial a_{i}\partial a_{j}} = -2\lambda_{4}\begin{pmatrix}
v^{2} & -v^{2}\\
-v^{2} & v^{2}
\end{pmatrix}.
\end{equation}
The actual mass values are the eigenvalues of these matrices:
\begin{equation}
m_{H}^{2} = M_{A}^{2} = \frac{-2\lambda_{4}}{2\lambda+\lambda_{4}}\mu^{2},\qquad
m_{h}^{2} = 2\mu^{2},\qquad
m_{G}^{2}=0,
\end{equation}
with one massless Goldstone boson and a pair of degenerate particles with opposite parity. Therefore, for vacuum stability,\footnote{Applying full 8-field 2HDM systems leads to the same values and constraints, but with a few extra massless Goldstone bosons that do not alter the conclusion.}
\begin{equation}
\lambda>0,\qquad 0>\lambda_{4}>-2\lambda.
\label{limit}
\end{equation}
As in ~\ref{app:2hdm}, we set $M_{h}=1$ and scale all parameters accordingly, and then use $\mu=0.5$, $\lambda=1.78$, $\lambda_{4}=-2.56$ in the analysis. The same parameter range is obtained from the 2HDM derivation.

\section{2HDM mass parameters}
\label{app:2hdm}

From previous research, the values of $\mu$ and $\lambda$ can be expressed in terms of the particle masses and mixing angles \cite{Battye_2020,PhysRevD.76.039902,Hespel_2014}:
 
\begin{equation}
\begin{aligned}
    \mu_{1} = &\frac{1}{2M_{h}^{2}}\bigl[M_{h}^{2}\cos^{2}a + M_{H}^{2}\sin^{2}a \\&\qquad+ (M_{h}^{2}-M_{H}^{2})\cos a\sin a\tan b\bigr],
\end{aligned}
\end{equation}
\begin{equation}
\begin{aligned}
\mu_{2} = &\frac{1}{2M_{h}^{2}}\bigl[M_{h}^{2}\cos^{2}a + M_{H}^{2}\sin^{2}a \\&\qquad+ (M_{h}^{2}-M_{H}^{2})\cos a\sin a\cot b\bigr],
\end{aligned}
\end{equation}
\begin{equation}
\lambda_{1} = \frac{M_{h}^{2}\cos^{2}a + M_{H}^{2}\sin^{2}a}{2M_{h}^{2}\cos^{2}b},
\end{equation}
\begin{equation}
\lambda_{2} = \frac{M_{h}^{2}\sin^{2}a + M_{H}^{2}\cos^{2}a}{2M_{h}^{2}\sin^{2}b},
\end{equation}
\begin{equation}
\lambda_{3} = \frac{(M_{h}^{2}-M_{H}^{2})\cos a\sin a + 2M_{H_{\pm}}^{2}\cos b\sin b}{M_{h}^{2}\cos b\sin b},
\end{equation}
\begin{equation}
\lambda_{4} = \frac{M_{A}^{2} - 2M_{H_{\pm}}^{2}}{M_{h}^{2}},
\end{equation}
\begin{equation}
\lambda_{5} = \frac{M_{A}^{2}}{M_{h}^{2}},
\end{equation}
  
normalized by the original Higgs boson mass $M_{h}=125\ \mathrm{GeV}$. The VEVs are given by \cite{Battye_2021}
\begin{equation}
v_{1} = \cos b\,v_{\text{SM}},\qquad v_{2} = \sin b\,v_{\text{SM}},
\end{equation}
with $v_{\text{SM}}=246\ \mathrm{GeV}$. Other values are set to $M_{H}=M_{A}=M_{H_{\pm}}=200\ \mathrm{GeV}$ \cite{Battye_2020} for simplicity. For results with other parameter choices, see the GitHub repository in ~\ref{app:git}.

In the early random trials of Sec.~\ref{sec:natural}, we used the mixing parameters $\tan a = \tan b = 0.85$ \cite{Battye_2020}. The highly radiation‑suppressed condition was found to require full mixing for $\tan a=\tan b=1$. The full mixing regime corresponds to equal VEVs for the two doublets and yields a symmetric potential that suppresses radiation modes. The parameters at this point are $\mu_{1}=\mu_{2}=0.5$, $\lambda_{1}=\lambda_{2}=1.78$, $\lambda_{3}=3.56$, $\lambda_{4}=-2.56$, $\lambda_{5}=2.56$.

All parameters are dimensionless after scaling by $M_{h}$. To convert back to SI units, multiply by the appropriate power of $M_{h}$.

In Sec.~\ref{sec:ideal}, the full-mixing condition $a=b=\pi/4$ is chosen because it exactly cancels the higher-harmonic radiation terms, as shown in equation ~(\ref{eq:cancellation}). This is not a fine-tuned point but rather a structural condition: whenever the Lagrangian parameters satisfy the relation $\lambda_1=\lambda_2=\lambda_3/2$ and $\lambda_4=-|\lambda_5|$, the cancellation holds. For the 2HDM mass parametrization, this corresponds to a one-dimensional surface in parameter space, not an isolated point. For example, taking $m_H=m_A=m_{H^\pm}=400\ \mathrm{GeV}$ with the same full-mixing condition yields an oscillon profile that follows the same trend as Fig.~\ref{f0f1} but exhibits slight quantitative deviations (e.g., a small shift in the peak radius and amplitude). These deviations do not affect the existence of the oscillon or the radiation-suppressed nature of the solution; they merely reflect the dependence of the profile on the mass scale. Thus, the radiation-suppressed oscillon exists on a continuous locus of the 2HDM parameter space; the specific masses chosen in the main text are for numerical convenience.

Table~\ref{tab:params} summarizes representative parameter points. The bias term $m_{12}^2$ is the soft $\mathbb{Z}_2$-breaking term added to equation (\ref{eq:potential}).\footnote{In the 2HDM potential, adding the term $m_{12}^2(\Psi_1^\dagger\Psi_2+\Psi_2^\dagger\Psi_1)$ can break the $\mathbb{Z}_2$ symmetry. When $m_{12}^2=0$, the two vacua are exactly degenerate (unbiased). A non-zero $m_{12}^2$ lifts this degeneracy, creating a bias between the vacua, and the domain wall network annihilates on a timescale set by $m_{12}^2$.} The rows with $\tan a=\tan b=1.0$ satisfy the full-mixing condition $(a=b=\pi/4)$, where higher-harmonic radiation is suppressed. The remaining rows demonstrate that the oscillon can still exist even when the ideal radiation-suppressed condition is not exactly met, though the lifetime is shorter. The structural condition for radiation suppression is $\tan a=\tan b=1$, regardless of the specific mass values, though, there is a range where mass values considered must make function parameters always satisfy the vacuum stability condition ($\lambda>0,\ 0>\lambda_4>-2\lambda$) given in  equation (~\ref{limit}).

\begin{table*}[!ht]
\centering
\small
\begin{tabular}{ccccc}
\hline\hline
$M_H,M_A,M_{H^\pm}$ (GeV) & $(\tan a,\tan b)$ & $m_{12}^2$ & Natural formation in random setups? & Ideal oscillon solution exists? \\
\hline
200 & (0.85, 0.85) & 0 & Yes & No (no convergent profile) \\
200 & (0.85, 1.0)  & 0 & Yes & No (no convergent profile) \\
200 & (1.0, 0.85)  & 0 & Yes & No (no convergent profile) \\
200 & (1.0, 1.0)   & 0 & Yes & Yes (radiation-suppressed) \\
200 & (1.2, 1.0)   & 0 & Yes & No (no convergent profile) \\
200 & (1.0, 1.2)   & 0 & Yes & No (no convergent profile) \\
200 & (1.2, 1.2)   & 0 & Yes & No (no convergent profile) \\
400 & (1.0, 1.0)   & 0 & Yes & Yes (radiation-suppressed) \\
400 & (0.85, 0.85) & 0.01 & Yes & No (no convergent profile) \\
400 & (1.0, 1.0)   & 0.01 & Yes & No (no convergent profile) \\
400 & (0.85, 0.85) & 0.05 & No & No (no convergent profile) \\
400 & (1.0, 1.0)   & 0.05 & No & No (no convergent profile) \\
\hline\hline
\end{tabular}
\caption{Representative parameter points for the 2HDM.  "Yes" in column 4 (natural formation) means the oscillon structure can be reproduced in the random initial condition simulations of Fig.~\ref{fig:domain_wall}; "Yes" in column 5 (ideal oscillon) means the profile can be constructed as in Fig.~\ref{f0f1} using the corresponding parameter set. $\tan a=\tan b=0.85$ corresponds to the random initial condition simulations in Sec.~\ref{sec:natural}. Rows with $\tan a=\tan b=1.0$ and $m_{12}^2=0$ satisfy the full-mixing condition where higher-harmonic radiation is suppressed. Other rows show that oscillons can still form naturally even away from the ideal point, though constructing the exact radiation-suppressed profile is not possible. The radiation-suppressed condition depends only on the mixing angles, not on the mass scale.}
\label{tab:params}
\end{table*}

\section{Floquet analysis details}
\label{app:floquet}

In the framework of Floquet theory, we investigate the linear stability by introducing a perturbation vector $\delta_{i}$ discretized on a radial grid with spacing $\Delta r$. The index $i$ runs over the $N_{f}$ field components ($N_{f}=4$ or $8$). The linearized equations of motion are
\begin{equation}
\ddot{\delta}_{i} = \hat{D}\delta_{i} + \hat{J}(t)_{ij}\delta_{j} - \hat{F}\dot{\delta}_{i},
\end{equation}
where $\hat{D}=\nabla^{2}$ is the spatial Laplacian, $\hat{J}(t)_{ij}=\frac{\partial^{2}V}{\partial\psi_{i}\partial\psi_{j}}$ is the Jacobian evaluated on the background trajectory, and $\hat{F}$ is an additional friction term that makes the sponge‑layer absorbing boundary condition consistent (if excluded, all Floquet multipliers lie on the unit circle). The radial domain $R$ is discretized into $N$ grid points, giving $2\times N_{f}\times N$ ODEs. For the 2HDM this results in $16N$ degrees of freedom. The Laplacian is represented by a tridiagonal matrix.

Define the state vector
\begin{equation}
\vec{v} = \bigl(\delta_{1}(0),\delta_{1}(\Delta r),\dots,\delta_{1}(R),\delta_{2}(0),\dots,\delta_{4}(R)\bigr)^{T},
\end{equation}
and $\vec{V}=(\vec{v},\dot{\vec{v}})^{T}$. Then
\begin{equation}
\begin{pmatrix}\dot{\vec{v}}\\ \ddot{\vec{v}}\end{pmatrix} = \begin{pmatrix}0 & \hat{\mathbb{I}}\\ \hat{D}+\hat{J} & -\hat{F}\end{pmatrix}\begin{pmatrix}\vec{v}\\ \dot{\vec{v}}\end{pmatrix},
\end{equation}
so that $\dot{\vec{V}} = A\vec{V}$. The Floquet matrix $M$, defined by $\vec{V}(T)=M\vec{V}(0)$ (where $T$ is the period), is given by
\begin{equation}
M = e^{\int_{0}^{T} A dt} = \exp\left(\int_{0}^{T}\begin{pmatrix}0 & \hat{\mathbb{I}}\\ \hat{D}+\hat{J} & -\hat{F}\end{pmatrix} dt\right).
\end{equation}
The integration is performed numerically with time steps $dt\ll T$. For stability, the moduli of the eigenvalues of $M$ must not exceed $1$. In the analysis of Sec.~\ref{sec:natural}, we used $dt=0.1$, $R_{\text{max}}=32$, and $\Delta r=0.25$~\footnote{To avoid orbit drift (accumulation of truncation errors that cause a computed periodic solution to deviate from its closed trajectory), one may need to increase the time resolution to $dt\sim10^{-5}T$; see the code for details.}.

\section{Numerical convergence}
\label{app:convergence}

To ensure that the observed long‑lived behavior is physical and not a numerical artifact, we perform a spatiotemporal convergence analysis. We adopt a representative non‑ideal oscillon configuration with $\omega=0.45$ in a potential tuned to exhibit non‑vanishing radiation:
\begin{equation}
\begin{aligned}
V(\Psi_{1},\Psi_{2}) &= -\mu^{2}(\Psi_{1}^{\dagger}\Psi_{1}) - \mu^{2}(\Psi_{2}^{\dagger}\Psi_{2}) \\
&\quad +0.8\lambda(\Psi_{1}^{\dagger}\Psi_{1})^{2} + 0.8\lambda(\Psi_{2}^{\dagger}\Psi_{2})^{2} \\
&\quad +2.4\lambda(\Psi_{1}^{\dagger}\Psi_{1})(\Psi_{2}^{\dagger}\Psi_{2}) + 2\lambda_{4}\left[\mathrm{Re}(\Psi_{1}^{\dagger}\Psi_{2})\right]^{2},
\end{aligned}
\end{equation}
with parameters from ~\ref{app:2hdm} with full mixing (all quantities are dimensionless).

\paragraph{Spatial convergence}
We vary the grid spacing $dx\in\{1.0,0.5,0.25\}$ while keeping $dt=0.1$ and the domain size $L=64$. Table~\ref{tab:convergence} shows that the total energy at $t=160$ converges as $dx$ decreases. The relative difference in $E(t=160)$ between $dx=0.5$ and $dx=0.25$ is about $2.8\%$, indicating that $\Delta r=0.25$ is sufficient for the main text. For efficiency we use $dx=0.5$ for time‑saving runs and $dx=0.25$ for higher accuracy.

\begin{table}[!t]
\centering
\begin{tabular}{lccc}
\hline\hline
Grid width $dx$ & $E(t=0)$ & $E(t=160)$ & $\Delta E$ (loss) \\
\hline
1.0   & 16.65 & 10.31 & 6.34 \\
0.5   & 17.24 & 11.01 & 6.23 \\
0.25  & 17.40 & 11.32 & 6.08 \\
\hline\hline
\end{tabular}
\caption{Spatial convergence test for a non‑ideal oscillon.}
\label{tab:convergence}
\end{table}

\paragraph{Temporal convergence}
We vary $dt\in\{0.1,0.05,0.025\}$ with $dx=0.25$. Table~\ref{tab:convergence_time} shows that the initial energy is invariant and the final energy varies by less than $0.3\%$, confirming that $dt=0.1$ satisfies the CFL condition and sufficiently resolves the high‑frequency oscillations.

\begin{table}[!t]
\centering
\begin{tabular}{lccc}
\hline\hline
Time step $dt$ & $E(t=0)$ & $E(t=160)$ & $\Delta E$ (loss) \\
\hline
0.1    & 17.34 & 11.32 & 6.02 \\
0.05   & 17.34 & 11.30 & 6.04 \\
0.025  & 17.34 & 11.29 & 6.05 \\
\hline\hline
\end{tabular}
\caption{Temporal convergence test.}
\label{tab:convergence_time}
\end{table}

\section{Data availability}
\label{app:git}

The source code and animations are available at \url{https://github.com/zhaoyu-meng/2HDM.git}. The repository includes:
1. GIF files for ideal structure animations,
2. C++ simulation code for 2D/3D lattice evolution,
3. Python scripts for Floquet analysis and $f_{0},f_{1}$ plotting.

\bibliography{bib}   

\end{document}